\pdfoutput=1
\documentclass[10pt,journal,cspaper,compsoc]{IEEEtran}

\usepackage{times}
\usepackage{epsfig}
\usepackage{graphicx}
\usepackage[cmex10]{amsmath} 
\usepackage{amssymb}
\usepackage[tight,normalsize,sf,SF]{subfigure}
\usepackage{epstopdf}
\usepackage{amsthm}
\usepackage{algorithm}
\usepackage{algorithmic}
\usepackage[nocompress]{cite}
\usepackage{url}
\usepackage{multirow}
\usepackage{tabularx}
\usepackage{enumerate}
\usepackage{cite}
\usepackage{color}
\newcommand{\citeauthor}[1]{{#1} \textit{et al.}}

\begin{document}
\title{Modularity in Complex Multilayer Networks with Multiple Aspects: A Static Perspective}

\author{Han~Zhang,~\IEEEmembership{Student Member,~IEEE,}
        Chang-Dong Wang,~\IEEEmembership{Member,~IEEE,}\protect\\
        Jian-Huang~Lai,~\IEEEmembership{Senior Member,~IEEE,}
        and Philip S. Yu,~\IEEEmembership{Fellow, IEEE}
\IEEEcompsocitemizethanks{
\IEEEcompsocthanksitem
Han Zhang, Chang-Dong Wang and Jian-Huang Lai are
with School of Data and Computer Science, Sun Yat-sen University, Guangzhou, P. R. China.\protect\\
E-mail: zh950713@gmail.com, changdongwang@hotmail.com, stsljh@mail.sysu.edu.cn.
\IEEEcompsocthanksitem
Philip S. Yu is with the University of Illinois at Chicago, Chicago, IL
60607, USA. \protect\\
E-mail: psyu@cs.uic.edu.}
\thanks{}}

\markboth{Journal of \LaTeX\ Class Files,~Vol.~6, No.~1, January~2007}%
{Shell \MakeLowercase{\textit{et al.}}: Bare Demo of IEEEtran.cls
for Computer Society Journals}

\IEEEcompsoctitleabstractindextext{%
\begin{abstract}
Complex systems are usually illustrated by networks which captures the topology of the interactions between the entities.
To better understand the roles played by the entities in the system one needs to uncover the underlying community structure of the system.
In recent years, systems with interactions that have various types or can change over time between the entities have attracted an increasing research attention.
However, algorithms aiming to solve the key problem --- community detection --- in multilayer networks are still limited.
In this work, we first introduce the multilayer network model representation with multiple aspects, which is flexible to a variety of networks.
Then based on this model, we naturally derive the multilayer modularity --- a widely adopted objective function of community detection in networks --- from a static perspective as an evaluation metric to evaluate the quality of the communities detected in multilayer networks.
It enables us to better understand the essence of the modularity by pointing out the specific kind of communities that will lead to a high modularity score.
We also propose a spectral method called mSpec for the optimization of the proposed modularity function based on the supra-adjacency representation of the multilayer networks.
Experiments on the electroencephalograph network and the comparison results on several empirical multilayer networks demonstrate the feasibility and reliable performance of the proposed method.
\end{abstract}

}

\maketitle

\IEEEdisplaynotcompsoctitleabstractindextext

\IEEEpeerreviewmaketitle

\section{Introduction}
\label{sec:introduction}
\IEEEPARstart{C}{omplex} systems are usually illustrated by networks which captures the topology of the interactions between the entities \cite{strogatz:2001exploring,newman:2010networks,wasserman:1994social,girvan:2002community,lambiotte2014random}.
For systems with more complicated entity interconnections, edges with different attributes, e.g. directed graphs\cite{newman:2010networks,bang:2008:digraphs}, weighted graphs\cite{newman:2004analysis,barrat:2004architecture}, signed graphs\cite{doreian:2009partitioning,yang:2007community} and so on, have been thoroughly studied.
In recent years, systems with entity interactions that have various types or can change over time have attracted an increasing research attention\cite{verbrugge:1979multiplexity,szell:2010multirelational,rocklin:2013clustering,holme:2012temporal}.
For example, a person interacts with his friends in Facebook and uses emails for business will demonstrate different behaviors in Facebook social network and email social network.
Such networks are usually interpreted as a combination of different ``layers" (or ``views", ``edge colors", ``relations", ``slices", \textit{etc.} in the literature), and is regarded as  \textit{multilayer networks}.
In different contexts, ``multigraph", ``multiplex network", ``multirelational network", ``multislice network", ``multilevel network", ``network of network" and ``temporal network" always refer to a similar network structure\cite{kivela:2014multilayer}.
Following the conventional terminology in network science, we refer to networks with such structure as \textbf{multilayer networks}.

 \begin{table*}[!htp]
\begin{center}
\caption{Notations used in this paper.}
\label{tab:notation}
\begin{tabular}{| c | c | c | c |}
\hline
\multicolumn{2}{|c|}{Single-layer Networks} &\multicolumn{2}{c|}{Multilayer Networks}\\
\hline
Notation &Description &Notation &Description\\
\hline
$N$ &The total number of nodes &$N$, $V_v$, $F$ &The total number of nodes within a layer, total layers within aspect $v$, and aspects \\
\hline
$A_{ij}$ &The adjacency matrix &$A_{ijs}^{\{v\}}$, $C_{ijs}^{\{v\}}$ &The within-layer and between-layer adjacency of layer $s$ in aspect $v$ (view $s^{\{v\}}$)\\
\hline
$B_{ij}$ &The modularity matrix &$B_{isjr}^{\{vw\}}$ &The multilayer modularity matrix \\
\hline
$p_{ij}$ &The adjacency of a null model &$p_{is}^{\{v\}}$ &The adjacency of a null model in layer $s^{\{v\}}$\\
\hline
$k_i$ &The node strength of node $i$ &$k_{is}^{\{v\}}$, $c_{is}^{\{v\}}$ &The within-layer and between-layer node strength of node $i$ of layer $s^{\{v\}}$\\
\hline
$m$ &The total number of edges &$m_{s}^{\{v\}}$ &The total number of edges within layer $s^{\{v\}}$\\
\hline
$g_i$ &The community label of node $i$ &$g_{is}^{\{v\}}$ &The community label of node $i$ of layer $s^{\{v\}}$\\
\hline
$Q$, $\mathcal{H}$ &The modularity and Hamiltonian &$Q_M$, $\mathcal{H}_M$ &The multilayer modularity and Hamiltonian \\
\hline
&&$\tilde{C}_{isr}^{\{vw\}}$ &The coupling strength of node $i$ between layer $s^{\{v\}}$ and layer $r^{\{w\}}$\\
\hline
&&$e_{isr}^{\{v\}}$, $\omega$ &The parameters to control the couplings between between layer $s^{\{v\}}$ and layer $r^{\{w\}}$\\
\hline
&&$\gamma_{s}^{\{v\}}$ &The resolution parameter controlling the contribution of the null model to the modularity\\
\hline
&&$\mu$, $\mu'$ &The normalization factor of multilayer modularity proposed in this paper and \cite{mucha:2010community}\\
\hline
\end{tabular}
\end{center}
\end{table*}


Although there is actually no consensus on its definition, a \textit{community} usually refers to a group of nodes that are compactly connected with each other and sparsely connected with those nodes outside the group.
By partitioning a network into communities we obtain its community structure, which is a coarse-grained representation of the network that assists us analyzing the roles played by each node\cite{fortunato:2010community}.
Despite numerous studies on multilayer networks in recent years, there is still a lack of evaluation metrics for measuring the community structure of a multilayer network, which in turn limits the number of available algorithms to find the optimal community structure in multilayer networks.
Existing evaluation metrics in multilayer networks are mainly derived from ``single-layer" cases, where the evaluation metrics are designed to detect modular structures in conventional networks that can be represented simply with nodes and edges, e.g. edge centrality, clustering coefficient, and metrics based on dynamic process\cite{battiston:2013metrics,brodka:2010method,de:2013centrality,lambiotte:2012ranking,kivela:2014multilayer,de2015:identifying}.
In such methods, detections are applied independently on each layers before final assignment, or on an ``collapsed network" which is a single-layer network generated by aggregating the layers\cite{peixoto:2015inferring}.
Such treatment is intuitive to find an ``average" role played by a node in different layers, but somehow fails to treat the multiple layers fundamentally as a whole.
In \cite{mucha:2010community}, \citeauthor{Mucha} proposed a modularity-based metric for multilayer network community structure derived from a Laplacian dynamic.
To the best of our knowledge, they for the first time introduce \textit{couplings} to the multilayer network models, which are links that appear between layers and connect a node with its copy in other layers, to combine the layers and form an interconnected-layer network model.
Based on such an interconnected-layer structure, the generalized modularity is able to evaluate the community structure without any compression or loss of the information encoded in the multilayer networks.

In spite of the great advances, the generalized modularity still has weaknesses which will lead to confusion especially when it comes to temporal networks, where the layers are usually time slices of a specific evolving single-layer network.
The derivation of modularity is based on the stability of global communities\cite{lambiotte:2008laplacian}, which is measured by comparing the position of a random walker with the stable state.
This assumes that the random walker keeps transferring as the time goes.
However, the layers are interdependent w.r.t. time and the couplings are introduced to describe the continuity of the interaction between nodes along the layers (time slices)\cite{mucha:2010community}.
It is confusing since one layer may be the result of the evolvement of another layer but the random walker is assumed to be able to travel between them.
Another important weakness is, although the generalized modularity is generally similar with its original version in the single-layer case, the definition of a community in multilayer networks becomes vaguer to understand --- what does a community look like in multilayer networks if a random walker can hardly escape from it?
Actually, the current derivation of multilayer modularity focus on capturing the dynamic property of a community --- stopping the random walker from leaving it.
In some cases where there is such random process defined on the network, the definition of the community is apparent.
But in other cases, the definition of a community becomes vague.

The above two issues are inevitably brought by the derivation from a dynamic perspective.
In order to address them, in this paper, we derive the generalized modularity from a static perspective, i.e. without defining dynamic process on the network.
As will be shown in section \ref{sec:ourmetric}, from such perspective, the generalized modularity is represented as the predominant part of \textit{Hamiltonian}, which measures the total energy of the systems in a variety of cases including community structure in the networks\cite{reichardt:2006statistical}.
Thus the optimization of the proposed metric is equivalent to that of Hamiltonian, which provides the generalized modularity with an energy explanation.
We also demonstrate in section \ref{sec:ourmetric} that the generalized modularity just finds communities with high cohesion, i.e. densely distributed internal efficient edges (not the couplings), which is more intuitive to understand and returns to its original definition in the single-layer case\cite{newman:2006modularity}.
With such a static derivation, we are able to generalize the modularity to multiple aspect cases, where the layers belong to different groups\cite{kivela:2014multilayer} or the layer relation is flexible.
We also propose a spectral algorithm called mSpec for optimizing the proposed modularity evaluation metric, which extends the spectral bisection algorithm in the single-layer case\cite{newman:2006modularity}.

We summarize our contribution in this paper briefly as follows:
\begin{itemize}
\item We derive the multilayer modularity from a static perspective to address the confusion in temporary networks and point out which kind of topological structure will lead to a high modularity value.
\item We generalize the multilayer modularity to adapt to networks with multiple aspects or there are flexible constraints on the layer relation.
\item We propose a spectral bisection algorithm (mSpec) for multilayer modularity optimization based on the supra-adjacency representation for multilayer structure.
\item We apply the proposed metric to electroencephalogram (EEG) networks as an attempt of application.
\end{itemize}

The rest of this paper will be arranged as follows. We review the related works that have been done in the literature in section \ref{sec:relatedwork}. The proposed multilayer modularity and the mSpec optimization will be described in section \ref{sec:ourmetric} and section \ref{sec:optimization} respectively. The experimental results are reported in section \ref{sec:experiment}. We conclude this paper in section \ref{sec:conclusion}.
For clarity, Table \ref{tab:notation} summarizes the notations used in this paper.

\section{Background}
\label{sec:relatedwork}
In this section, we will briefly introduce the network models that have been explored in the literature and the strategies that have been adopted to detect communities in multilayer networks, including evaluation metrics and optimization.
\subsection{Network Model}
\label{sec:bakgndmodel}
\begin{figure}[!htp]
\centerline{
\includegraphics[scale=0.35]{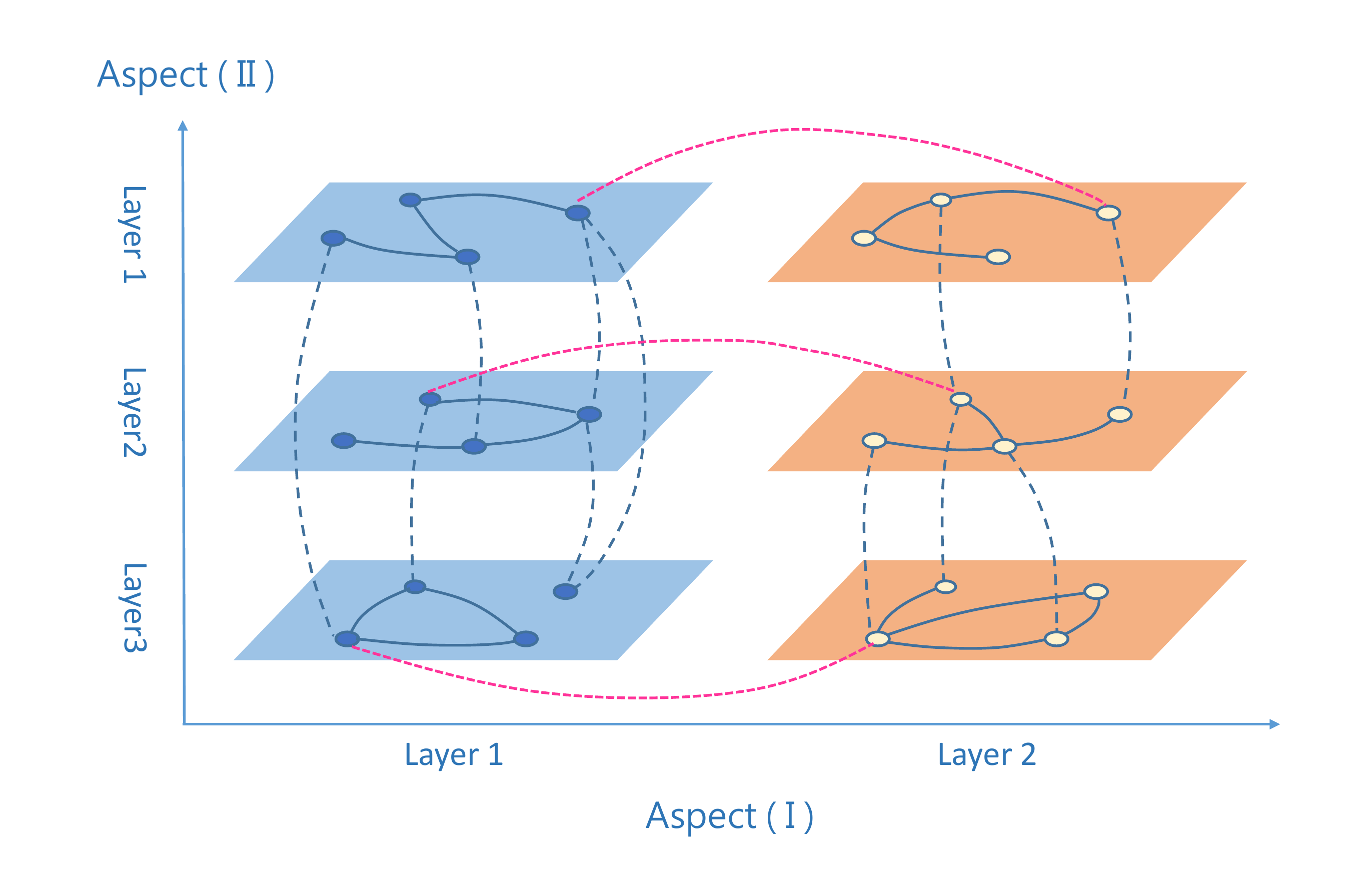}
}
\caption{Aspect-aspect representation of the multilayer network model\cite{kivela:2014multilayer}.
Aspect (I) has two layer sets and aspect (II) has three layer sets.
The within-layer edges are denoted with solid lines and the couplings are denoted with dotted lines where different colors indicate the couplings of different aspects.
}
\label{fig:multiviewmodel}
\end{figure}
During the process of exploring the multilayer networks, different network models have been proposed\cite{mucha:2010community,boccaletti:2014structure,kivela:2014multilayer}.
\citeauthor{Mucha} linked multiple single-layer networks with \textit{couplings}, which refer to the edges that connect the nodes with their copies in different layers, to represent a multilayer network\cite{mucha:2010community}.
This model allows the layers to communicate through the couplings and is widely adopted especially by research involving dynamics defined on multilayer networks\cite{de:2013mathematical,gomez:2013diffusion}.
\citeauthor{De Domenico} proposed a multilayer model based on tensor representation, which no longer restrains the between-layer connection to appear between node-copy pairs\cite{de:2013mathematical}.
In the rest of this paper, we will use \textit{between-layer edges} to refer to this kind of connections that link a node with another node in different layers.
On the one hand, the presence of between-layer edges makes the network more flexible.
But on the other hand, a multilayer network with between-layer edges is very similar to a single-layer network in structure since they both have no limitations on the presence of edges (any node in any layer is allowed to link with another node in another layer).
This will sometimes blur the boundary between single-layer and multilayer networks.

In more complex systems, the layers may be divided into several groups, which indicates that multilayer networks should also be distinguished when observed from different \textit{aspects}\cite{kivela:2014multilayer}.
For example, a cellphone contact network can be characterized by different means such as calling and texting.
Meanwhile, this network is also temporal since there are callings and textings at any time point.
Thus this layers is divided into two groups: according to time stamps or according to communication means.
In order not to lose the information of the networks from either aspect, we have to construct a more complex multilayer network.
There are actually two types of multilayer networks with multiple aspects, which has not been clearly distinguished in the literature.
If a layer can belong to more than one aspects, which means the aspects may overlap, we can locate a single layer by indicating all aspects it belongs to.
In the rest of this paper, we will call this an \textit{aspect-aspect} representation, as shown in \figurename~\ref{fig:multiviewmodel}.
In such representation, we need a $F$-dimensional (the number of aspects) vector to locate a layer.
For example the layer at the top right can be located by (2, 1) since the layer is in layer set 2 of aspect (I) and layer set 1 of aspect (II).
When there are additional aspects, the dimension of the location vector grows.
Therefore, it is a challenging task to represent this network by matrix with predetermined size.

\begin{figure}
\centerline{
\includegraphics[scale=0.35]{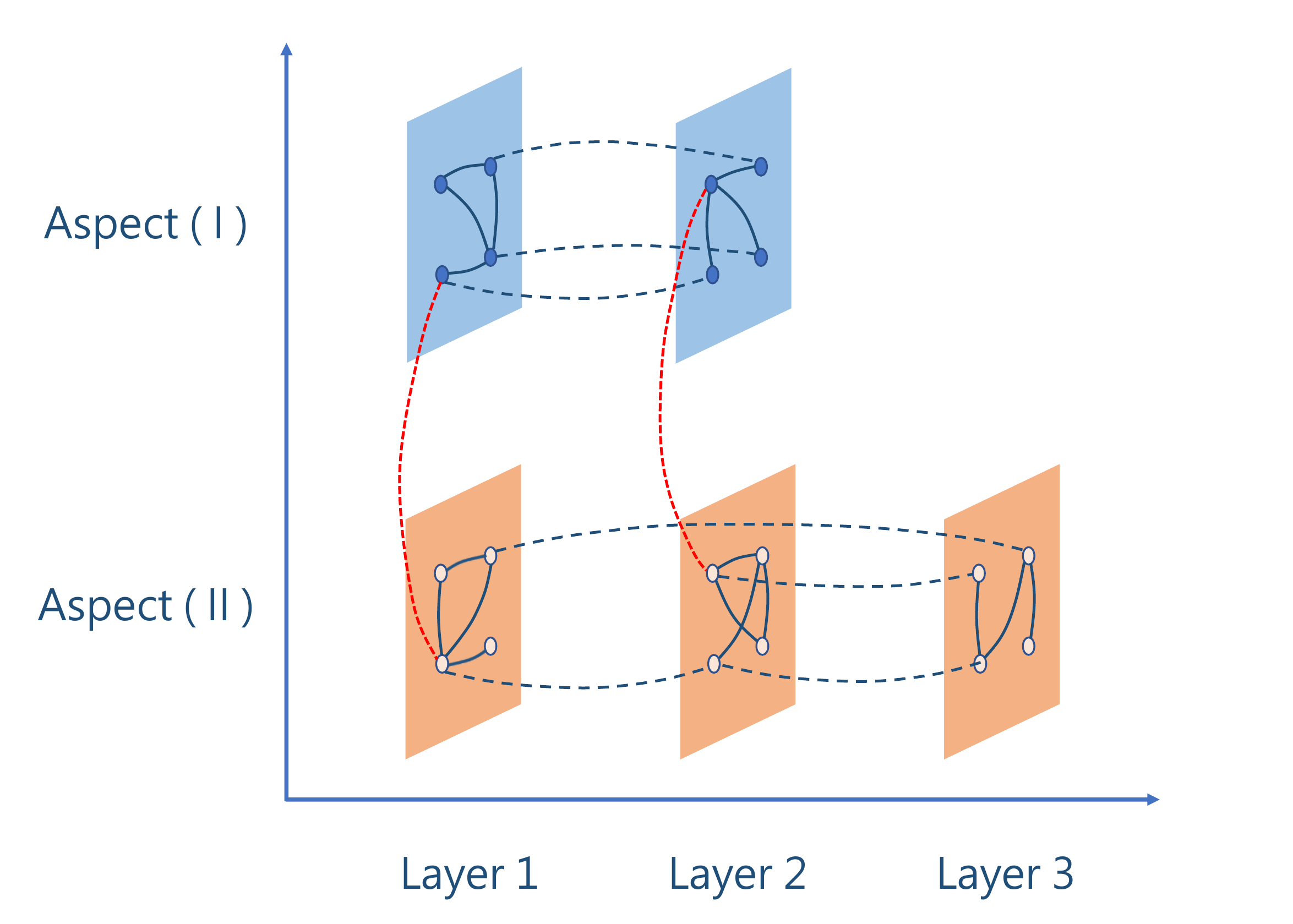}
}
\caption{Aspect-layer representation of the multilayer network model.
The within-layer edges are denoted with solid lines and the couplings are denoted with dotted lines where different colors indicate the couplings of different aspects.
The position of a layer in a multilayer network can be specified by determining which aspect it belongs to and its serial number within the aspect.
For example, the layer at the bottom right can be located as $(2, 3)$ since it is in aspect (II) and layer 3.
Such representation avoid the problem caused by increasing aspect numbers.
}
\label{fig:ourmodel}
\end{figure}

In other networks with multiple aspects, each layer only belongs to a unique aspect.
For instance, conventional electroencephalogram (EEG) networks (single-layer networks) for different individuals construct a multilayer network, where each layer corresponds to an EEG network of a person.
With the fact that all testees receive the same treatment, the layers reflect the common reactions to the test, but still hold the individual difference between the testees.
In addition, a person can take several EEG tests to obtain different EEG networks that all reflect the roles played by different regions of his brain in the test.
Thus we have two aspects observing the EEG network of the testees, enabling us to analyze individual differences and similarities as well as the role of different brain regions simultaneously.
With respect to different individuals, we may have as many aspects as the number of persons that takes the EEG test, and the layers within that aspect are several EEG network obtained from several tests.
To locate a layer in such networks, we just need to point out to which aspect the layer belongs and its position within that aspect, as shown in \figurename~\ref{fig:ourmodel}.
We will refer to such representation as \textit{aspect-layer} representation to distinguish with the aspect-aspect representation.

Actually, for a more convenient implementation, we can convert the aspect-aspect representation to the aspect-layer representation by absorbing aspects hierarchically into one aspect.
We can interpret this process by considering how multidimensional arrays are stored on the disk.
A 2-dimensional array is represented as an ``array of arrays".
The multiple aspects are arranged in a similar way so that we can represent the network using matrices with predetermined size.

\subsection{Existing Evaluation Metrics for Community Detection in Multilayer Networks}
\label{sec:relatedmetric}
As one of the most concerned issues in network analysis, \textit{community detection} aims at partitioning the network into groups of closely connected nodes (which is called a \textit{community}) to obtain a coarse-grained representation, which helps us better understand the structure of the network.
However, as far as we are concerned, most of existing evaluation metrics designed for community detection in multilayer networks assume that the layers are independent.
The multilayer stochastic blockmodels (SBM), which are generative models that make inferences on the role of nodes given the network structure as evidence\cite{valles:2014multilayer,peixoto:2015inferring}, usually adopt two types of strategies.
They either learn a SBM on each layer, just like in single-layer networks, and then make global assignments based on the result of each layers, or they aggregate the layers to produce a ``collapsed network"\cite{peixoto:2015inferring}.
The final community assignment of each node is made based on the SBM result on the collapsed network.

\citeauthor{De Domenico} extended the well-known \textit{infomap} method\cite{rosvall:2008maps} to the multilayer case\cite{de2015:identifying}.
The infomap method solves the community detection problem by considering its duality with a coding problem.
It assumes that the community is able to capture the flows on the network so that by utilizing the community structure, we can greatly compress the coding length needed to describe a random process on the network\cite{rosvall:2008maps}.
The goal is to minimize the ``map equation", which describes the coding length based on a specific partition and the transition probability of the random process.
\citeauthor{De Domenico} defined the transition probability of a random walker in multilayer networks so that the map equation is able to describe the flow in multilayer scenarios.
Such treatment is intuitively correct, albeit they assume the node can reach the neighbors of its copies in other layers in a single step.
In fact this implicitly erases the difference between layers --- it is equivalent to consider a collapsed network.

Some other existing evaluation metrics also provide considerable solutions to the community detection problem in multilayer networks, such as multilayer clustering coefficient (the authors consider the overlapping of layers or the networks with multiple types of connections)\cite{brodka:2010method,battiston:2013metrics}, multilayer centrality (the authors consider a random walker to jump between layers through specific node pairs or edges)\cite{de:2013centrality,lambiotte:2012ranking}, \textit{etc}.
What these methods share in common is that they assume the layers are independent or can be aggregated and attempt to find global roles for the nodes.
Such treatments would have considerable effects as the network structure varies when we wish to find the similarity of the layers.
But when we are interested in the different roles of nodes in the layers, these methods may generate a poor result, as we will discuss in the experiments.
Thus it is highly recommended to adopt an interconnected-layer structure.

\textit{Modularity} is a widely adopted metric for community detection in single-layer networks\cite{newman:2004finding,newman:2010networks,clauset:2004finding,newman:2006modularity}.
The original definition of modularity is the edge difference between the current network and a \textit{null model}, which is a rewired random network with the same degree distribution as the original network.
Modularity reflects the cohesion of nodes within a community, so by optimizing global modularity one can find a partition of the network with communities within which the edges are densely distributed\cite{newman:2006modularity}.
Recently, \citeauthor{Mucha} extended the single-layer modularity to multilayer case using a Laplacian dynamic process defined on the multilayer network (without between-layer edges), which measures the stability of a community by comparing the probability of a random walker to stay in the same community at time $t$ to the static solution (i.e. $t \rightarrow \infty$) \cite{mucha:2010community,lambiotte:2008laplacian}.

This generalized modularity is of great contribution due to the fact that it combines the layers (using the couplings) on a model level for the first time and is adopted in a wide range of areas\cite{szell:2010multirelational,bassett:2011dynamic,chiu:2011unifying}.
Nevertheless, this evaluation metric still has weaknesses.
The multilayer modularity is derived based on a dynamic process (actually it is a random walk process), which means the random walker is jumping between nodes as time goes.
So what if the network is evolving over time?
When it comes to temporal networks, whose layers can be interpreted as different time slices of an evolving network (i.e. the edges varies over time), things get confusing, because the layers can be seen as different states in the network evolving process.
Moreover, although the within-layer representation is the same as the conventional form proposed by \citeauthor{Newman}, it is not clear what kind of community the multilayer modularity tends to find.
It is of vital importance to know the bias of the evaluation metrics on the communities, so that we can pick appropriate evaluation metrics for corresponding network structures.
Last but not the least, the coupling strength strategy needs modification to adapt to more general cases, since the original one is brought without much discussion.

\subsection{Optimization}
Optimizing the single-layer modularity is an NP-hard problem\cite{brandes:2008modularity}, so we can only obtain a good approximation of the optimal solution efficiently.
Since the single-layer modularity is actually a component of the multilayer modularity, the optimization of the multilayer modularity will also be NP-hard.
To our best knowledge, there are rare algorithms except a generalized Louvain heuristic approach for multilayer modularity optimization\cite{mucha:2010community}.
The Louvain method is a greedy iterative method which hierarchically aggregates two nodes into a group by making the optimal modularity gain in each iteration.
Then the generated node group is regarded as a new node and another iteration starts.
This algorithm converges when there is no such merger that increases global modularity value.
Some tricks like adding a Kernighan-Lin node swapping step\cite{kernighan:1970efficient} after each iteration will give better detection result.
The Louvain method is a widely adopted heuristic for optimizing quality functions of community structure, which implies that it does not utilize the property of the evaluation metric.
Meanwhile, the community assignments of nodes are not guaranteed to converge to a good approximation, so we may need to run the algorithm several times to obtain a relatively more reasonable solution.
As will be discussed in section \ref{sec:experiment}, we cannot control the community scale detected by the Louvain method.
When it comes to EEG networks, the Louvain method provides a relatively fine-grained detection result, whereas we expect it to find two communities --- the regions that are active or inactive.

In order to tackle the above issues, we adopt the aspect-layer representation for describing network structure which is intuitive to implement and derive the multilayer modularity from a static perspective (not involving the dynamic process).
We also discuss the extension of the evaluation metric so as to make it applicable when considering different types of multilayer networks such as unbalanced multilayer networks, temporal networks or signed networks \textit{etc.}
We propose a spectral method for optimizing the multilayer modularity which provides a stable solution and is helpful when we concern the scale of the discovered communities.

\section{Multilayer Modularity from a Static Perspective}
\label{sec:ourmetric}
We start with several general requirements that a quality function should satisfy as introduced in \cite{reichardt:2006statistical}: (i) rewarding existing edges within a community, (ii) penalizing non-existing edges within a community, (iii) penalizing existing edges between two communities and (iv) rewarding non-existing edges between two communities. Thus a general quality function takes the form
\begin{equation}
\label{eq:single-layeransatz}
\begin{aligned}
&\mathcal{H}(g) = -\sum_{i\neq{j}}a_{ij}\underbrace{A_{ij}\delta(g_i, g_j)}_{\text{Internal existing edges}}+ \sum_{i\neq{j}}b_{ij}\underbrace{(1-A_{ij})\delta(g_i, g_j)}_{\text{Internal non-existing edges}}\\
&+\sum_{i\neq{j}}c_{ij}\underbrace{A_{ij}\big[1-\delta(g_i, g_j)\big]}_{\text{External existing edges}}-\sum_{i\neq{j}}d_{ij}\underbrace{(1-A_{ij})\big[1-\delta(g_i, g_j)\big]}_{\text{External non-existing edges}},
\end{aligned}
\end{equation}
where $A_{ij}$ is the edge strength of nodes $i$ and $j$, $g_i$ indicates the label of the community that node $i$ belongs to, and $a$, $b$, $c$, $d$ are free parameters. The delta function $\delta$ is the Kronecker delta.
In multilayer networks, since there are three kinds of edges (within-layer edges, couplings and between-layer edges), we need to expand this function to enable the additional edge types. To be more explicit, the between-layer edges will be ignored in this paper since they blur the boundaries between such multilayer model and a single-layer network (i.e. both of them have no restraints on the appearance of the edges). But similar tricks can be designed to easily enable between-layer edges in this model. The expanded quality function can be written as
\begin{equation}
\label{eq:multilayeransatz}
\begin{aligned}
&\mathcal{H}_{M}(g)= -\sum_{i\neq{j}}\sum_{v=1}^{F}\sum_{s=1}^{V_v}\underbrace{a_{ijs}^{\{v\}}A_{ijs}^{\{v\}}\delta(g_{is}^{\{v\}}, g_{js}^{\{v\}})}_{\text{Within-layer internal existing edges}}\\
&~~+\sum_{i\neq{j}}\sum_{v=1}^{F}\sum_{s=1}^{V_v}\underbrace{b_{ijs}^{\{v\}}(1-A_{ijs}^{\{v\}})\delta(g_{is}^{\{v\}}, g_{js}^{\{v\}})}_{\text{Within-layer internal non-existing links}} \\
&~~+\sum_{i\neq{j}}\sum_{v=1}^{F}\sum_{s=1}^{V_v}\underbrace{c_{ijs}^{\{v\}}A_{ijs}^{\{v\}}[1-\delta(g_{is}^{\{v\}}, g_{js}^{\{v\}})]}_{\text{Within-layer external existing links}} \\
&~~-\sum_{i\neq{j}}\sum_{v=1}^{F}\sum_{s=1}^{V_v}\underbrace{d_{ijs}^{\{v\}}(1-A_{ijs}^{\{v\}})[1-\delta(g_{is}^{\{v\}}, g_{js}^{\{v\}})]}_{\text{Within-layer external non-existing links}} \\
&~~-\sum_{sv\neq{rw}}\sum_{i=1}^{N}\underbrace{e_{isr}^{\{vw\}}C_{isr}^{\{vw\}}\delta(g_{is}^{\{v\}}, g_{ir}^{\{w\}})}_{\text{Between-layer internal existingcouplings}} \\
&~~+\sum_{sv\neq{rw}}\sum_{i=1}^{N}\underbrace{f_{isr}^{\{vw\}}(1-C_{isr}^{\{vw\}})\delta(g_{is}^{\{v\}}, g_{ir}^{\{w\}})}_{\text{Between-layer internal non-existing couplings}} \\
&~~+\sum_{sv\neq{rw}}\sum_{i=1}^{N}\underbrace{g_{isr}^{\{vw\}}C_{isr}^{\{vw\}}[1-\delta(g_{is}^{\{v\}}, g_{ir}^{\{w\}})]}_{\text{Between-layer external existing couplings}} \\
&~~-\sum_{sv\neq{rw}}\sum_{i=1}^{N}\underbrace{h_{isr}^{\{vw\}}(1-C_{isr}^{\{vw\}})[1-\delta(g_{is}^{\{v\}}, g_{ir}^{\{w\}})]}_{\text{Between-layer external non-existing couplings}},
\end{aligned}
\end{equation}
where we use $s$ and $r$ for the denotation of different layers, $v$ and $w$ for that of aspects. $N$ and $V_v$ represent the total number of nodes within a layer and total number of layers of aspect $v$ and matrix $\mathbf{A}$, $\mathbf{C}$ and $\mathbf{g}$ denote the within-layer adjacency, between-layer adjacency and the community label matrix, respectively. The number of parameters has doubled after taking between-layer couplings into account.
Eq. \eqref{eq:single-layeransatz} points out the general form of an objective function for community detection, and can be used to derive the \textit{Hamiltonian} of a Potts model in statistical mechanics as well as the modularity\cite{wu:1982potts,reichardt:2006statistical}, while Eq. \eqref{eq:multilayeransatz} restricts the quality that an objective function in the multilayer case should satisfy.
Since the parameters of Eq. \eqref{eq:multilayeransatz} control the punishment (encouragement) and are free to choose, we can take $a_{ijs}^{\{v\}}=c_{ijs}^{\{v\}} = 1-b_{ijs}^{\{v\}} = 1-d_{ijs}^{\{v\}} = 1-\gamma_{s}^{\{v\}}p_{ijs}^{\{v\}}$ and $e_{isr}^{\{vw\}} = f_{isr}^{\{vw\}} = g_{isr}^{\{vw\}} = h_{isr}^{\{vw\}}$ to obtain a similar representation as the multilayer modularity\cite{mucha:2010community}, where $p_{ijs}^{\{v\}}$ known as null model is the penalty factor, and the parameter $\gamma_s^{\{v\}}$ known as resolution parameter balances the contribution of punishment and award.
Thus we obtain a Hamiltonian function
\begin{equation}
\label{eq:multilayerHamiltonianOriginal}
\begin{aligned}
&\mathcal{H}_M(g) = -\sum_{ijsv}(A_{ijs}^{\{v\}}-\gamma_{s}^{\{v\}}p_{ijs}^{\{v\}})[2\delta(g_{is}^{\{v\}}, g_{js}^{\{v\}})-1] \\
&~~-\sum_{isrvw}e_{isr}^{\{vw\}}[2\delta(g_{is}^{\{v\}}, g_{ir}^{\{w\}})-1](2C_{isr}^{\{vw\}}-1).
\end{aligned}
\end{equation}
In Eq. \eqref{eq:multilayerHamiltonianOriginal}, we notice that the terms that do not contain $\delta$ will be constant in optimization process, so we can rewrite $\mathcal{H}_M(g)$ as
\begin{equation}
\begin{aligned}
&\mathcal{H}_M(g) = -2\sum_{ijsv}(A_{ijs}^{\{v\}}-\gamma_{s}^{\{v\}}p_{ijs}^{\{v\}})\delta(g_{is}^{\{v\}}, g_{js}^{\{v\}}) \\
&~~-2\sum_{isrvw}e_{isr}^{\{vw\}}(2C_{isr}^{\{vw\}}-1)\delta(g_{is}, g_{ir}) \\
&~~+\sum_{sv}(1-\gamma_{s}^{\{v\}})\cdot{2m_{s}^{\{v\}}} + \sum_{isrvw}(2C_{isr}^{\{vw\}}-1)e_{isr}^{\{vw\}}.
\end{aligned}
\end{equation}
By using $\tilde{C}_{isr}^{\{vw\}}=e_{isr}^{\{vw\}}(2C_{isr}^{\{vw\}}-1)$, we can get the standard Hamiltonian form for system with many particles
\begin{equation}
\label{eq:multilayerhamiltonian}
\begin{aligned}
&-\frac{1}{2}\mathcal{H}_{M}(g)= \\
&\sum_{isjrvw}\bigg[(A_{ijs}^{\{v\}}-\gamma_{s}^{\{v\}}p_{ijs}^{\{v\}})\delta_{sr}\delta_{vw}+\tilde{C}_{isr}^{\{vw\}}\delta_{ij}\bigg]\delta(g_{is}^{\{v\}}, g_{jr}^{\{w\}})\\
&-\frac{1}{2}\bigg\{\sum_{sv}(1-\gamma_{s}^{\{v\}})\cdot{2m_{s}^{\{v\}}}+ \sum_{isrw}\tilde{C}_{isr}^{\{vw\}}\bigg\},
\end{aligned}
\end{equation}
where the first term is proportional to Mucha's modularity\cite{mucha:2010community} (except the value $\tilde{C}_{isr}^{\{vw\}}$ takes differs from $C_{ijs}$ in \cite{mucha:2010community} which will be discussed in section \ref{sec:connection}).
The last two terms which can be interpreted as bias are dependent on the given network and the parameters.
Therefore, minimizing Hamiltonian is equivalent to optimizing modularity. Finally we obtain the modularity representation
\begin{equation}
\begin{aligned}
&Q_M(g) =\\
&\sum_{isjrvw}\bigg[(A_{ijs}^{\{v\}}-\gamma_{s}^{\{v\}}p_{ijs}^{\{v\}})\delta_{sr}\delta_{vw}+\tilde{C}_{isr}^{\{vw\}}\delta_{ij}\bigg]\delta(g_{is}^{\{v\}}, g_{jr}^{\{w\}}).
\end{aligned}
\end{equation}
Here $p_{ijs}^{\{v\}}$ is the within-layer edge strength of the null model. We can take different null models for different network types such as directed networks, and bipartite networks, \textit{etc.}\cite{mucha:2010community,bazzi:2014community}. Traditionally, in an undirected network we take Newman-Girvan null model (i.e. a uniform network) $\frac{k_{is}^{\{v\}}k_{js}^{\{v\}}}{2m_{s}^{\{v\}}}$ and this leads to
\begin{equation}
\begin{aligned}
\label{eq:ourmetric}
&~~~~~Q_M(g) =\\
&\sum_{isjrvw}\bigg[(A_{ijs}^{\{v\}}-\gamma_{s}^{\{v\}}\frac{k_{is}^{\{v\}}k_{js}^{\{v\}}}{2m_{s}^{\{v\}}})\delta_{sr}\delta_{vw}
+\tilde{C}_{isr}^{\{vw\}}\delta_{ij}\bigg]\delta(g_{is}^{\{v\}}, g_{jr}^{\{w\}}).
\end{aligned}
\end{equation}

Now we can take a closer look at the choice of the parameters in Eq. \eqref{eq:multilayeransatz}.
we take
\begin{equation}
  \begin{cases}
    a_{ijs}^{\{v\}}=c_{ijs}^{\{v\}} = 1-\gamma_{s}^{\{v\}}p_{ijs}^{\{v\}}\\
    b_{ijs}^{\{v\}} = d_{ijs}^{\{v\}} = \gamma_{s}^{\{v\}}p_{ijs}^{\{v\}}\\
    e_{isr}^{\{vw\}} = f_{isr}^{\{vw\}} = g_{isr}^{\{vw\}} = h_{isr}^{\{vw\}},
  \end{cases}
\end{equation}
which groups the edges into two types and gives different punishment (encouragement).
The value of the parameters are actually the efficient number of each type of edges (the edge difference of current network structure and the null model).
In other words, a positive modularity is obtained if the edges and couplings within the community are more than those between different communities.
A higher modularity is reached when the edges are more densely distributed within the communities.

\subsection{Selection of $\tilde{C}_{isr}^{\{vw\}}$}
\label{sec:connection}
In \cite{mucha:2010community}, \citeauthor{Mucha} proposed a multilayer modularity based on a Laplacian dynamic defined on multilayer network model as follows
\begin{equation}
\label{muchamodularity}
Q' = \frac{1}{2\mu'}\bigg[(A_{ijs}-\gamma_s\frac{k_{is}k_{jr}}{2m_s})\delta_{sr}-C_{jsr}\delta_{ij}\bigg]\delta(g_{is}, g_{jr}).
\end{equation}
Although similar to the proposed form as Eq. \eqref{eq:ourmetric} in structure (taking $v = w$ to obtain a single-aspect representation of modularity in this paper), \citeauthor{Mucha} did not discuss much about the coupling strength $C_{jsr}$.
They chose $C_{jsr}$ to take binary value $\{0, \omega\}$ to represent the absence and presence of couplings and $\omega$ controls the contribution of couplings.
In the proposed form, we notice that $\tilde{C}_{isr}^{\{vw\}} = e_{isr}^{\{vw\}}\cdot(2C_{isr}^{\{vw\}}-1)$ and if we take $e_{isr}^{\{vw\}} = \omega$, $\tilde{C}_{isr}^{\{vw\}}$ takes $\{-\omega, \omega\}$ representing the absence and presence of couplings in a specific community.
Compared with Mucha's modularity, the proposed form will punish those couplings that do not show up, so the couplings that are absent will also provide information about the community structure.
Additionally, since $\tilde{C}_{isr}^{\{vw\}} = e_{isr}^{\{vw\}}\cdot(2C_{isr}^{\{vw\}}-1)$ and $e_{isr}^{\{vw\}}$ is totally free, the proposed form of multilayer modularity is flexible to adjust to various types of multilayer networks.
We will use two typical types of network as an example.

\subsubsection{Unevenly Distributed Views}
Consider a common type of multilayer network whose distribution of layers is uneven, i.e. the intervals between pairs of layers can be unequal. In this situation, simply letting $\tilde{C}_{isr}^{\{vw\}}$ take the same value without considering the closeness of layers will cause large errors. For instance, suppose we have a multilayer electroencephalogram network in which each person is treated as a layer. Apparently, the age difference and gender of the patients will greatly influence the result\cite{sharma:2015developmental,repovs:2011brain}. Therefore, we should enable the proposed model to handle such networks with unevenly distributed layers. Noticing $e_{isr}^{\{vw\}}$ is a free parameter governing the amplitude of $\tilde{C}_{isr}^{\{vw\}}$, we can adjust $e_{isr}^{\{vw\}}$ according to the closeness of the layers as
\begin{equation}
e_{isr}^{\{vw\}} = \omega\cdot\frac{M_{sr}^{\{vw\}}}{\max_{s, r, v, w}{M_{sr}^{\{vw\}}}},
\end{equation}
where $M_{sr}$ measures the closeness of layer $s$ and $r$. Here we still use $\omega$ to control the coupling strength so as to control the balance between within-layer edges and between-layer couplings.

\subsubsection{Temporal Networks}
In some research a \textit{temporal network} is defined as a sequence of networks corresponding to successive time points with between-layer couplings indicating the continuity between adjacent layers\cite{holme:2012temporal,bazzi:2014community,berlingerio:2013multidimensional}. For example, suppose in a phone calling temporal network, two nodes are linked by an edge in two successive layers. If there is a coupling connecting the corresponding nodes in both layers, then we can tell that this call lasts through these two time points. Otherwise we can tell that they have two calls at both of the time points.
Therefore, between-layer couplings only appear between adjacent layers in such temporal networks.
In order to satisfy this we let  $e_{isr}^{\{vw\}} = 0$ when $|s-r| \neq 1$ or the link between the nodes does not last between two time points.

Notice that the interval between two time slices can also be unequal.
For example, the Facebook social networks of a person when he was 15 and 16 will be similar but they may have large difference compared with the network when he was 20.
Such time interval problem can be addressed just like the unevenly distributed layers discussed before.

\subsection{Signed Networks}
Connections in complex systems reflect either positive or negative interactions between nodes, which can be modeled as signed networks that contain edges with positive or negative weight\cite{doreian:2009partitioning,yang:2007community}.
The effect of both kinds of edges on the structure of such networks should be distinguished: the contribution of positive edges should be awarded, while the contribution of the negative edges should be punished.
In \cite{mucha:2010community}, Mucha \textit{et al.} derive the modularity by using a Laplacian dynamics operator that contains the sign information.
We can bring in signed edges into the proposed metric by representing the adjacency $A_{ijs}^{\{v\}}$ as well as the null model $p_{ijs}^{\{v\}}$ as the combination of both kinds of edges in Eq. \eqref{eq:multilayeransatz}
$$
A_{ijs}^{\{v\}} = A_{ijs}^{\{v\}+}-A_{ijs}^{\{v\}-},
$$
$$
\gamma_s^{\{v\}}p_{ijs}^{\{v\}} = \gamma_s^{\{v\}+}p_{ijs}^{\{v\}+}-\gamma_s^{\{v\}-}p_{ijs}^{\{v\}+}
$$
Thus we obtain the signed version of the proposed metric
\begin{equation}
\begin{aligned}
Q_M(g) = \frac{1}{\mu}&\sum_{ijsr}\bigg\{\bigg[\bigg(A_{ijs}^{\{v\}+}-\gamma_s^{\{v\}+}\frac{k_{is}^{\{v\}+}k_{js}^{\{v\}+}}{2m_s^{\{v\}+}}\bigg)\\
&-\bigg(A_{ijs}^{\{v\}-}-\gamma_s^{\{v\}-}\frac{k_{is}^{\{v\}-}k_{js}^{\{v\}-}}{2m_s^{\{v\}-}}\bigg)\bigg]\delta_{sr}\delta_{vw}\\
&+\tilde{C}_{isr}^{\{vw\}}\delta_{ij}\bigg\}\delta(g_{is}^{\{v\}}, g_{jr}^{\{w\}}).
\end{aligned}
\end{equation}
This is equivalent to considering the within-layer modularity as the combination of two ``networks" with opposite contribution. We can now conclude that the proposed metric is able to deal with signed networks by considering the negative edges as additional networks of the within-layer modularity.

\section{mSpec: An Iterative Spectral Optimization of Multilayer Modularity}
\label{sec:optimization}
In order to find a good approximation of the optimal solution of multilayer modularity maximization problem, \citeauthor{Mucha} adopted a generalized Louvain method, which hierarchically merges two communities to increase the modularity score\cite{mucha:2010community}.
The result is improved by a KL-swap step that swaps the nodes between the communities to see if further increase on modularity score is possible\cite{kernighan:1970efficient}.
But such optimization method is unstable, so we need to run it multiple times to avoid converging to a local maxima.
And it sometimes fails to find expected number (always small) of communities, since the algorithm stops before the number of communities decreases to the desired value.
\citeauthor{Newman} proposed a spectral method for single-layer modularity optimization which hierarchically divides the network into two communities\cite{newman:2006modularity}.
Inspired by their work, we propose a spectral bisection method called \textit{mSpec} based on the \textit{supra-adjacency} representation of the multilayer network.
This method will provide more stable performance as will be discussed in the experiments.

\subsection{Supra-adjacency representation: An Equivalent Single-layer Network}
\begin{figure}
\centerline{
\includegraphics[scale=0.48]{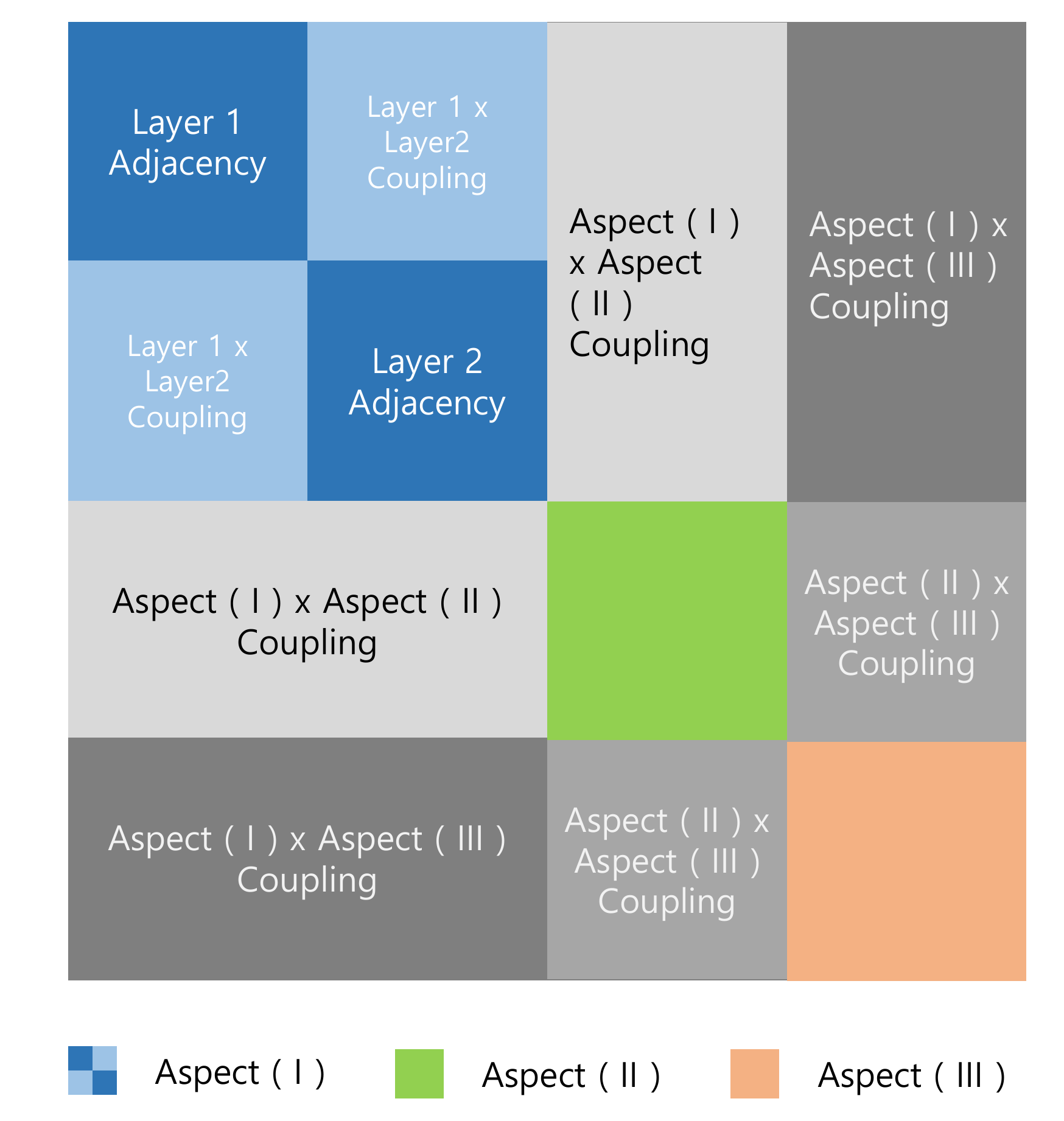}
}
\caption{Supra-adjacency matrix of a multilayer network with 3 aspects. The first aspect consists of two layers and the others contain only one layer. The non-diagonal blocks of the supra-adjacency matrix represent the between-layer adjacency of the layers. Since we only consider the between-layer couplings, these blocks are all diagonal. The diagonal blocks record the within-layer adjacency. Since we only discuss about undirected networks, the supra-adjacency matrix is symmetric.}
\label{fig:supranetwork}
\end{figure}

In multilayer network analysis, a \textit{supra-adjacency} always refers to a single-layer network which is flattened from a multilayer network\cite{kivela:2014multilayer,boccaletti:2014structure,sanchez:2014dimensionality,bazzi:2014community,cozzo:2015multilayer}. The basic idea is to combine two layers which are represented by two $N\times{N}$ graphs, to obtain an expanded layer which is represented by a $2\times{2}$ block graph with the diagonal blocks representing the within-layer adjacency of each layer and off-diagonal blocks representing the between-layer couplings. By repeating such flattening step until the number of layers reduces to one, we obtain an expanded equivalent single-layer network containing all nodes in the original multilayer network, where the nodes are distinguished from their copies in different layers and aspects (see Fig. \ref{fig:supranetwork}).

Based on the supra-adjacency representation, we obtain a mapping from a multilayer network to an equivalent single-layer network where the mapped subscript for node $i$ in layer $s^{\{v\}}$ is
\begin{equation}
\label{eq:mapping}
x = i + (s-1)N + \sum_{v'}^{v-1}V_{v'}N
\end{equation}
with $x  \in \big[1, \sum_{v}^FV_vN\big]$.

We therefore can apply the same mapping on the \textit{modularity matrix} which records the modularity of each node pair $(i,j)$ in each layer pair $(s^{\{v\}},r^{\{w\}})$ to obtain a \textit{supra-modularity-matrix}
\begin{equation}
\label{eq:modularity_matrix}
\begin{aligned}
B_{isjr}^{\{vw\}} &= \lambda_{s}^{\{v\}}\cdot\Big(A_{ijs}^{\{v\}}-\gamma_{s}^{\{v\}}\frac{k_{is}^{\{v\}}k_{js}^{\{v\}}}{2m_{s}^{\{v\}}}\Big)\delta_{sr}\delta_{vw} + \delta_{ij}\tilde{C}_{jsr}^{\{vw\}} \\
&= D_{xy}
\end{aligned}
\end{equation}
We will illustrate this supra-modularity-matrix maintains all the information of the original multilayer network and can be utilized for optimization.

\subsection{Dividing Networks into Two Communities}
Let the \textit{index matrix} $\mathcal{L}$ identify the community label of each node in each layer
\begin{eqnarray}
\mathcal{L}_{is}^{\{v\}}=
\begin{cases}
+1 & \text{if node $i$ in layer $s^{\{v\}}$ is in community 1}\\
-1 & \text{otherwise.}
\end{cases}
\end{eqnarray}
Then we can rewrite the modularity function as
\begin{equation}
Q = \frac{1}{\mu}\sum_{isjrvw}B_{isjr}^{\{vw\}}\Big(\frac{\mathcal{L}_{is}^{\{v\}}\mathcal{L}_{jr}^{\{w\}}+1}{2}\Big).
\end{equation}
We notice that
\begin{equation}
\begin{aligned}
\sum_{isjrvw}&B_{isjr}^{\{vw\}}\\
&= \lambda_{s}^{\{v\}}\cdot\sum_{ijs}(A_{ijs}^{\{v\}}-\gamma_{s}^{\{v\}}\frac{k_{is}^{\{v\}}k_{js}^{\{v\}}}{2m_{s}^{\{v\}}})+\sum_{isrvw}\tilde{C}_{isr}^{\{vw\}}\\
&= \sum_{sv}(1-\gamma_{s}^{\{v\}})\cdot{\lambda_{s}^{\{v\}}2m_{s}^{\{v\}}} + \sum_{isrvw}\tilde{C}_{isr}^{\{vw\}} \\
&= \chi
\end{aligned}
\end{equation}
which means once the graph is given, $\chi$ is a constant value and will not influence the global maximization of modularity function. Also, the $\frac{1}{\mu}$ and $\frac{1}{2}$ values in the parentheses do not make sense in the maximization, either. So our objective function can be rewritten as
\begin{equation}
\label{eq:QB}
Q = \sum_{isjrvw}B_{isjr}^{\{vw\}}\mathcal{L}_{is}^{\{v\}}\mathcal{L}_{jr}^{\{w\}}.
\end{equation}

Then we can map the multilayer network to the corresponding supra-adjacency as described in Eq. \eqref{eq:modularity_matrix}
\begin{equation}
\label{ass_modularity}
B_{isjr}^{\{vw\}} = D_{xy},
\end{equation}
where the mapping is performed according to Eq. \eqref{eq:mapping}. We can also bring in a new label vector \textbf{z} with
\begin{equation}
z_x = \mathcal{L}_{is}^{\{v\}}.
\end{equation}
Therefore, we can represent the objective function Eq. \eqref{eq:QB} as
\begin{equation}
\label{eq:QD1}
Q = \sum_{xy}D_{xy}z_xz_y.
\end{equation}
By applying this mapping, the problem is converted to be a relatively simple one, on which we can apply the same spectral method used in the single-layer case. We can solve it by utilizing the eigenvectors and eigenvalues of matrix $\mathbf{D}$ as follows
\begin{equation}
\label{eq:QD2}
Q = \sum_{xy}D_{xy}z_xz_y
  = \mathbf{z}^T\mathbf{Dz}.
\end{equation}
We can then represent \textbf{z} as the linear combination of the eigenvectors of $\mathbf{D}$, i.e. $\mathbf{z} = \sum_xa_x\mathbf{u}_x$, where $\mathbf{u}_x$ is the $x$-th eigenvector of $\mathbf{D}$ and $a_x$ is the corresponding weight. We can obtain $a_x = \mathbf{z}\cdot \mathbf{u}^T_{x}$. Meanwhile, if $\beta_x$ is the corresponding eigenvalue of $\mathbf{u}_x$, we can obtain $\mathbf{u}_x^T\mathbf{D} = (\mathbf{D}\cdot{\mathbf{u}_x})^T = \beta_x\mathbf{u}_x^T$ according to the fact that $\mathbf{D}$ is symmetric because $B_{jris}^{\{wv\}} = B_{isjr}^{\{vw\}}$ which means $D_{xy} = D_{yx}$. Then Eq. \eqref{eq:QD2} can be written as
\begin{equation}
\begin{aligned}
Q &= \sum_xa_x\mathbf{u}_x^T\mathbf{Dz} \\
  &= \sum_xa_x\mathbf{u}_x^T\cdot\mathbf{z}\beta_x \\
  &= \sum_xa_x^2\beta_x.
\end{aligned}
\end{equation}
We know that in order to maximize $Q$, supposing that the eigenvector corresponding to the largest eigenvalue is $\mathbf{u}_M$, all we need to do is to assign the vector \textbf{z} according to $\mathbf{u}_M$
\begin{equation}
\label{assignment}
z_x = \mathcal{L}_{is}^{\{v\}} =
\begin{cases}
1 &\ \text{if}\ [\mathbf{u}_M]_x \geq 0,\\
-1 &\ \text{otherwise}
\end{cases}
\end{equation}
Thus we obtain the optimal division using the supra-modularity-matrix.

\subsection{Dividing Networks into More than Two Communities}
To divide the network into more communities, we have to rewrite the additional modularity contribution of further division. Suppose the subcommunities after dividing community $C$ are $\mathcal{A}$ and $\mathcal{B}$, we have
\begin{equation}
\begin{aligned}
Q'_C &= Q_{\mathcal{A}} + Q_{\mathcal{B}} \\
&= \frac{1}{\mu}\bigg(\sum_{isv, jrw\in{\mathcal{A}}}B_{isjr}^{\{vw\}} + \sum_{isv, jrw\in{\mathcal{B}}}B_{isjr}^{\{vw\}}\bigg) \\
&= \frac{1}{\mu}\sum_{isv, jrw\in{C}}\frac{\mathcal{L'}_{is}^{\{v\}}\mathcal{L'}_{jr}^{\{w\}} + 1}{2} B_{isjr}^{\{vw\}},
\end{aligned}
\end{equation}
where $\mathcal{L'}_{jr}^{\{w\}} \in{\{-1, +1\}}$ is the community label indicating to $\mathcal{A}$ or $\mathcal{B}$ the node belongs. Here we use the fact that the sum of entries of modularity matrix $\mathbf{B}$ is constant once the network is determined so that it will not influence the optimization. Then the multilayer modularity gain can be written as
\begin{equation}
\begin{aligned}
&\Delta Q \\
&= Q'_{C} - Q_{C} \\
&=\frac{1}{2\mu}\bigg[\sum_{isv, jrw\in{C}}B_{isjr}^{\{vw\}}\mathcal{L}_{is}^{\{v\}}\mathcal{L}_{jr}^{\{w\}}-\sum_{isv, jrw\in{C}}B_{isjr}^{\{vw\}}\bigg]\\
&=\frac{1}{2\mu}\sum_{isv, jrw\in{C}}\bigg[B_{isjr}^{\{vw\}}-\delta_{ij}\delta_{sr}\delta_{vw}\sum_{j'r'w'\in{C}}B_{isj'r'}^{\{vw'\}}\bigg]\mathcal{L}_{is}^{\{v\}}\mathcal{L}_{jr}^{\{w\}}\\
&=\frac{1}{2\mu}\sum_{isv, jrw\in{C}}B_{isjr}^{\{vw\}(C)}\mathcal{L}_{is}^{\{v\}}\mathcal{L}_{jr}^{\{w\}}
\end{aligned}
\end{equation}
where each entry of matrix $\mathbf{B}^{(C)}$ is
\begin{equation}
B_{isjr}^{\{vw\}(C)} = B_{isjr}^{\{vw\}}-\delta_{ij}\delta_{sr}\delta_{vw}\sum_{j'r'w'\in{C}}B_{isj'r'}^{\{vw'\}}.
\end{equation}
Similarly, we also bring in an assistant matrix $\mathbf{D}$ to maximize the global modularity, $B_{isjr}^{\{vw\}} = D_{xy}$. Notice that $\sum_{j'r'w'\in{C}}B_{isj'r'}^{\{vw'\}}$ is constant, $\mathbf{B}^{(C)}$ is also symmetric and $\sum_{isjrvw}B_{isjr}^{\{vw\}(C )} = 0$, so we can repeatedly apply the bisection method on the detected communities using $\mathbf{D}$ as the modularity matrix $\mathbf{B}$ until the modularity gain $\Delta Q$ does not increase.

\subsection{Complexity Analysis}
The mSpec method is based on a linear mapping and spectral decomposition.
The time complexity of the linear mapping is $O(\sum_{v}^FV_vN)$, where $N$ is the total number of nodes in a single layer and $V_{v}$ is total layers within aspect $v$.
By applying Lanczos algorithm\cite{freund:1993implementation}, finding the dominant eigenvector can be done in $O((\sum_{v}^FV_vN)^2)$~\cite{newman:2006modularity}.
Thus, suppose there are $k$ divisions, we can complete the total calculation in time $O(k(\sum_{v}^FV_vN)^2)$.
The total number of divisions depends on the depth of the division tree, which is expected to be $\log(\sum_v^FV_vN)$ in average.
Thus the total complexity is $O(\log(\sum_v^FV_vN)(\sum_{v}^FV_vN)^2)$.
In summary, the theoretical efficiency of the mSpec method can be guaranteed as the scale of data grows.

\section{Experiments}
\label{sec:experiment}
In this section we present community detection results using the proposed modularity in several multilayer networks.
As we will demonstrate in the results, 1) the proposed method can be applied to a wide range of networks by flexibly adjusting the couplings and parameters; 2) the mSpec is more stable than the generalized Louvain method.

We conduct several experiments on a well-known benchmark network to discuss how the parameters can influence the results of community detection.
The proposed method is also applied to the electroencephalograph (EEG) networks as an attempt of its application, the result of which turns out to coincide with the functional division of the human brains.
In order to evaluate the performance of the proposed modularity optimization method (mSpec), it is compared with baseline optimization methods.
As will be reported, the proposed optimization performs more reliably as the coupling scale varies.

The networks we use in experiments are
\begin{enumerate}
\item \textbf{Parameter analysis data:}
\begin{itemize}
\item Zachary Karate Club network: network of friendships between 34 members of a karate club in a US university\cite{zachary:1977information}.
\end{itemize}
\item \textbf{Comparison data:}
\begin{itemize}
\item CKM-Physicians Innovation multilayer network: a network of the physicians' adoption of a new drug, tetracycline, in four towns\cite{coleman:1957diffusion}. There are 246 nodes and 3 layers (according to three questions asking about the relationship between the physicians).
\item CS-Aarhus social network: a multilayer social network consists of five online and offline relationships (5 layers) between 61 employees of Computer Science department at Aarhus\cite{magnani:2013combinatorial}.
\item Kapferer Tailor Shop network: a time-varying network recording the interactions in a tailor shop in Zambia over ten months\cite{kapferer:1972strategy}. The network consists of two layers according to the interaction types and 39 nodes.
\item Krackhardt High-Tech network: three kinds of social relationships (Advice, Friendship and ``Reports to") between 21 managers of a high-tech company\cite{krackhardt:1987cognitive}.
\item London Transportation network: multilayer transportation network of 369 London train stations with 3 layers recording different types of connection (Underground, Overground and DLR)\cite{de:2014navigability}. This network is relatively sparse.
\item Padgett Florentine Families network: the network of marriage alliances and business relationships between Florentine families in the Renaissance\cite{padgett:1993robust}. There are 16 nodes in total.
\item Vickers Class Relation network: the networks collected from 29 seventh grade students in an Australia school about three questions on the classmate relationship (``Get on with", ``Best friend" and ``Prefer to work with")\cite{vickers:1981representing}.
\end{itemize}
\item \textbf{Case Study data: EEG network}
\begin{itemize}
\item Signed multilayer network that characterize the correlation of the testees' brain regions during a visual stimuli test. The nodes include 128 scalp electrodes as well as a standard control electrode and 11 testees and several test records form a two-aspect multilayer network.
\end{itemize}
\end{enumerate}

\begin{figure*}[!htp]
\centerline{
\subfigure[$\omega = 0$]{\includegraphics[width=0.33\linewidth]{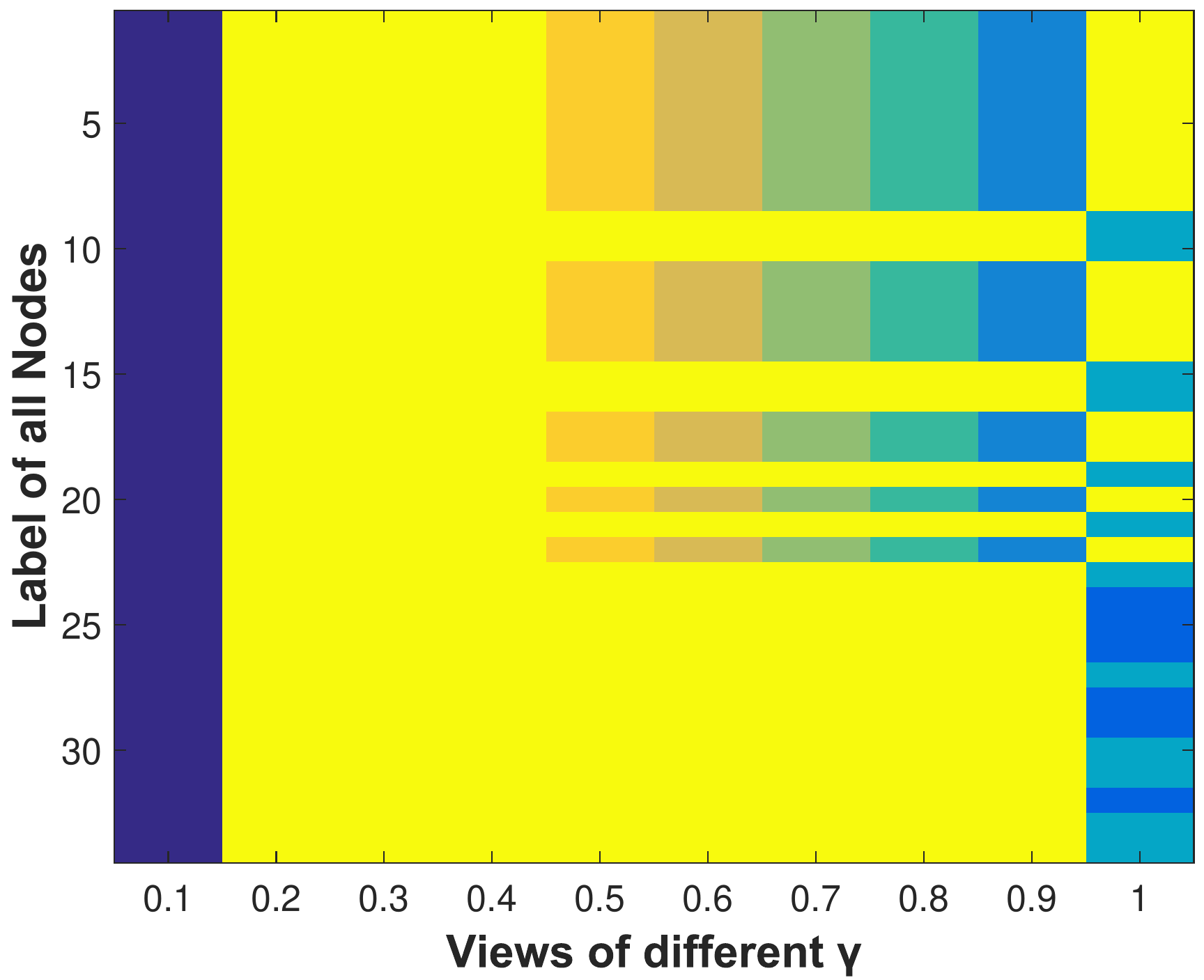}\label{fig:omega0}}%
\subfigure[$\omega = 0.01$]{\includegraphics[width=0.33\linewidth]{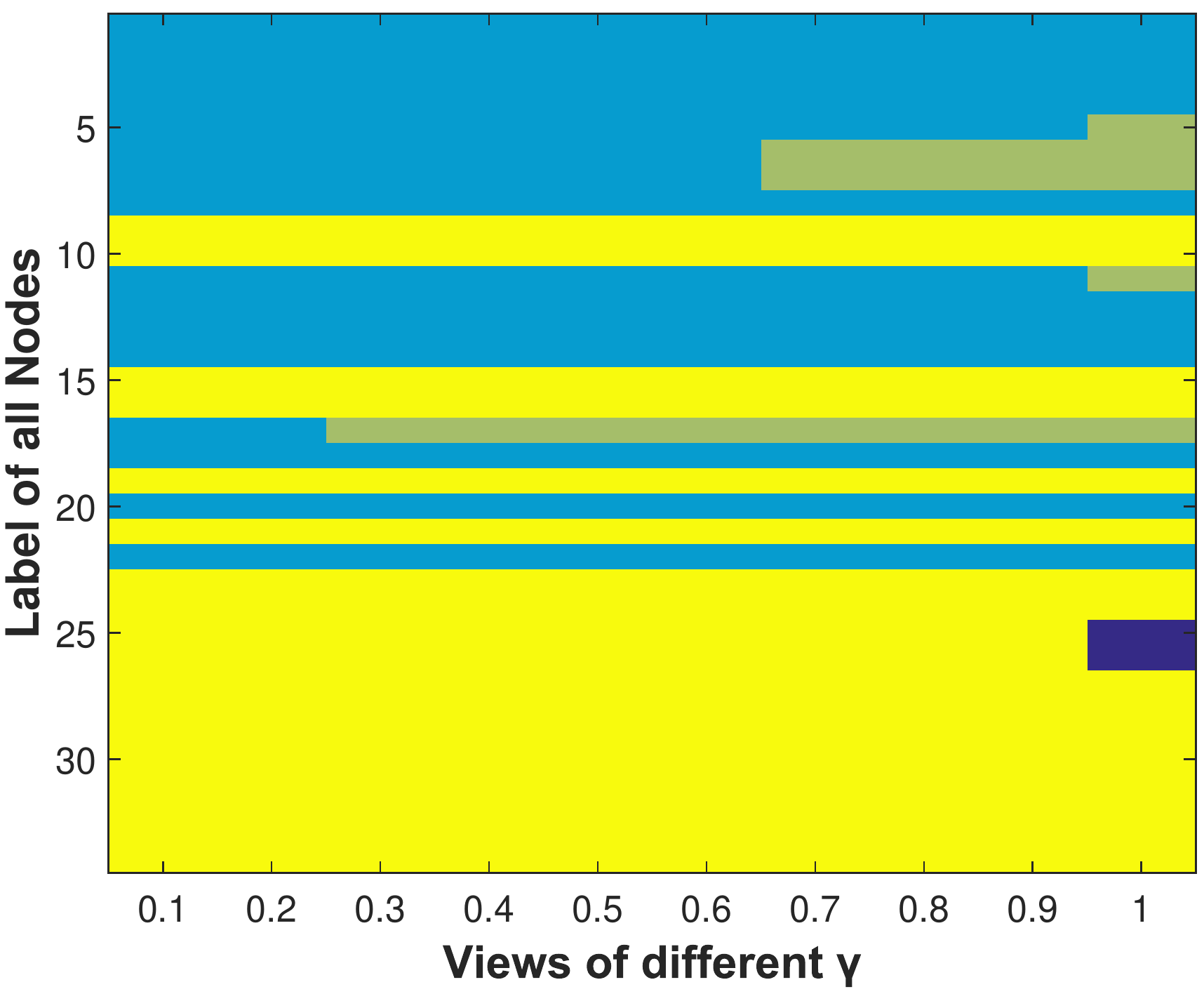}\label{fig:omega0.01}}%
\subfigure[$\omega = 0.1$]{\includegraphics[width=0.33\linewidth]{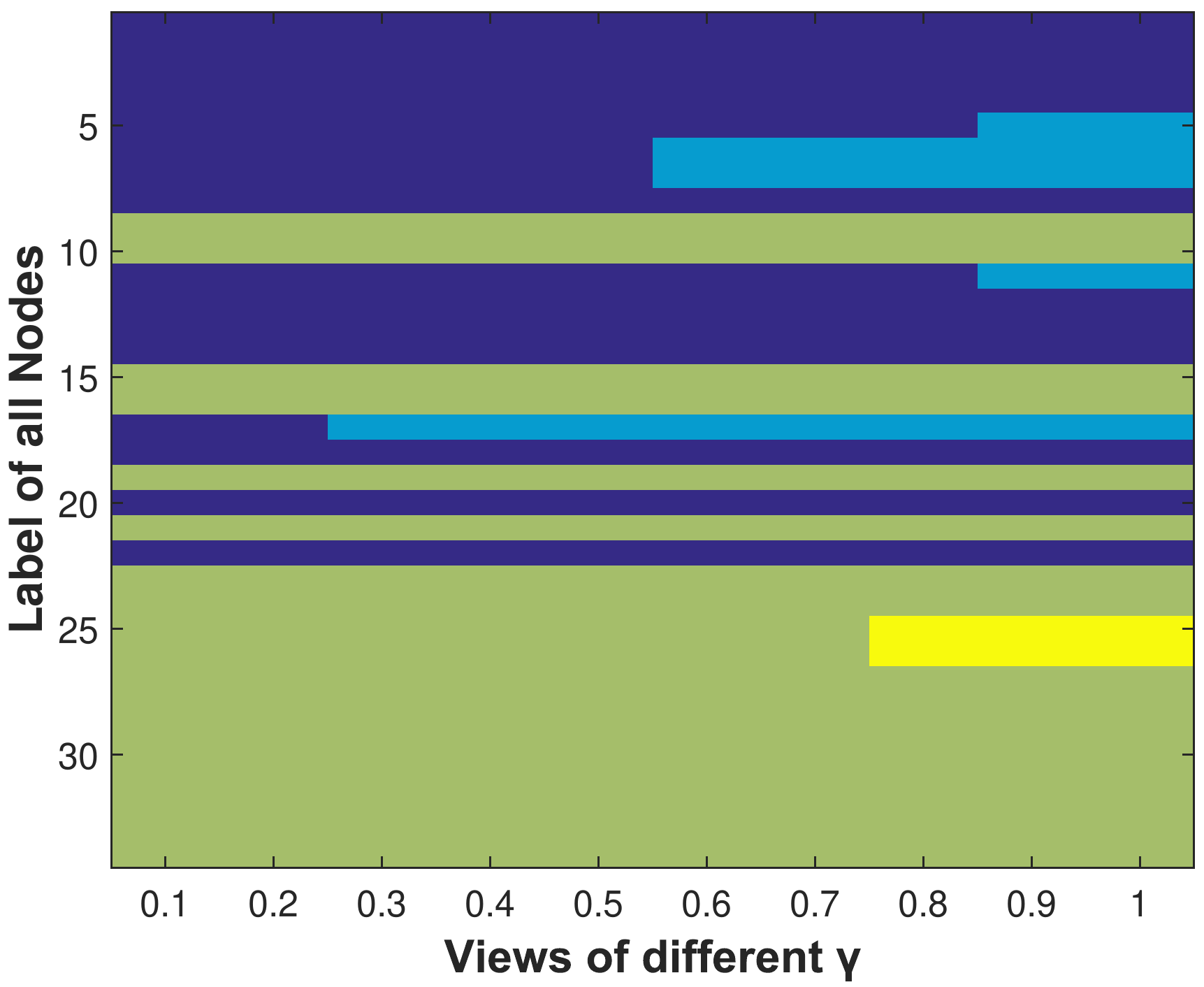}\label{fig:omega0.1}}%
}
\centerline{
\subfigure[$\omega = 1$]{\includegraphics[width=0.33\linewidth]{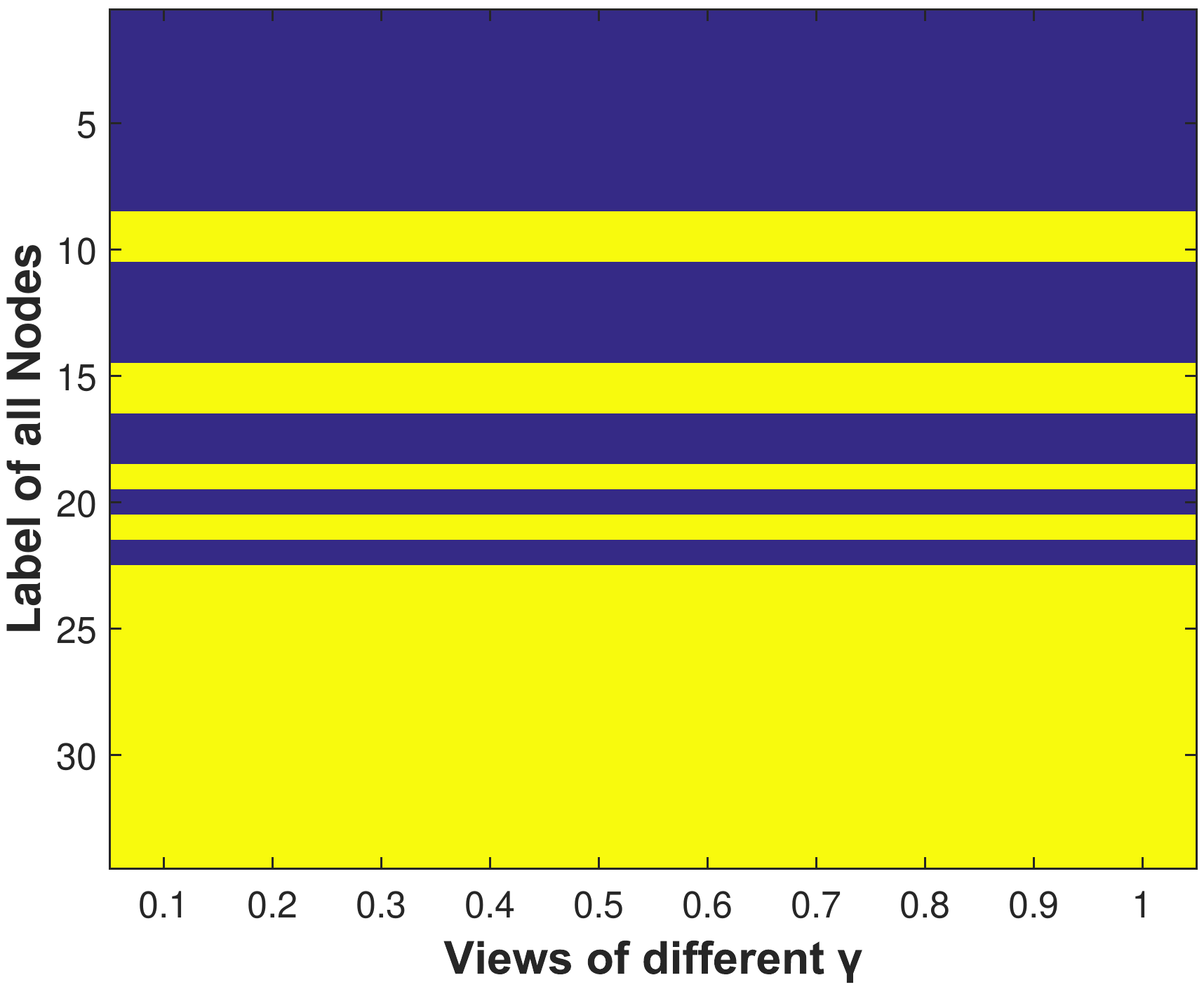}\label{fig:omega1}}%
\subfigure[$\omega = 10$]{\includegraphics[width=0.33\linewidth]{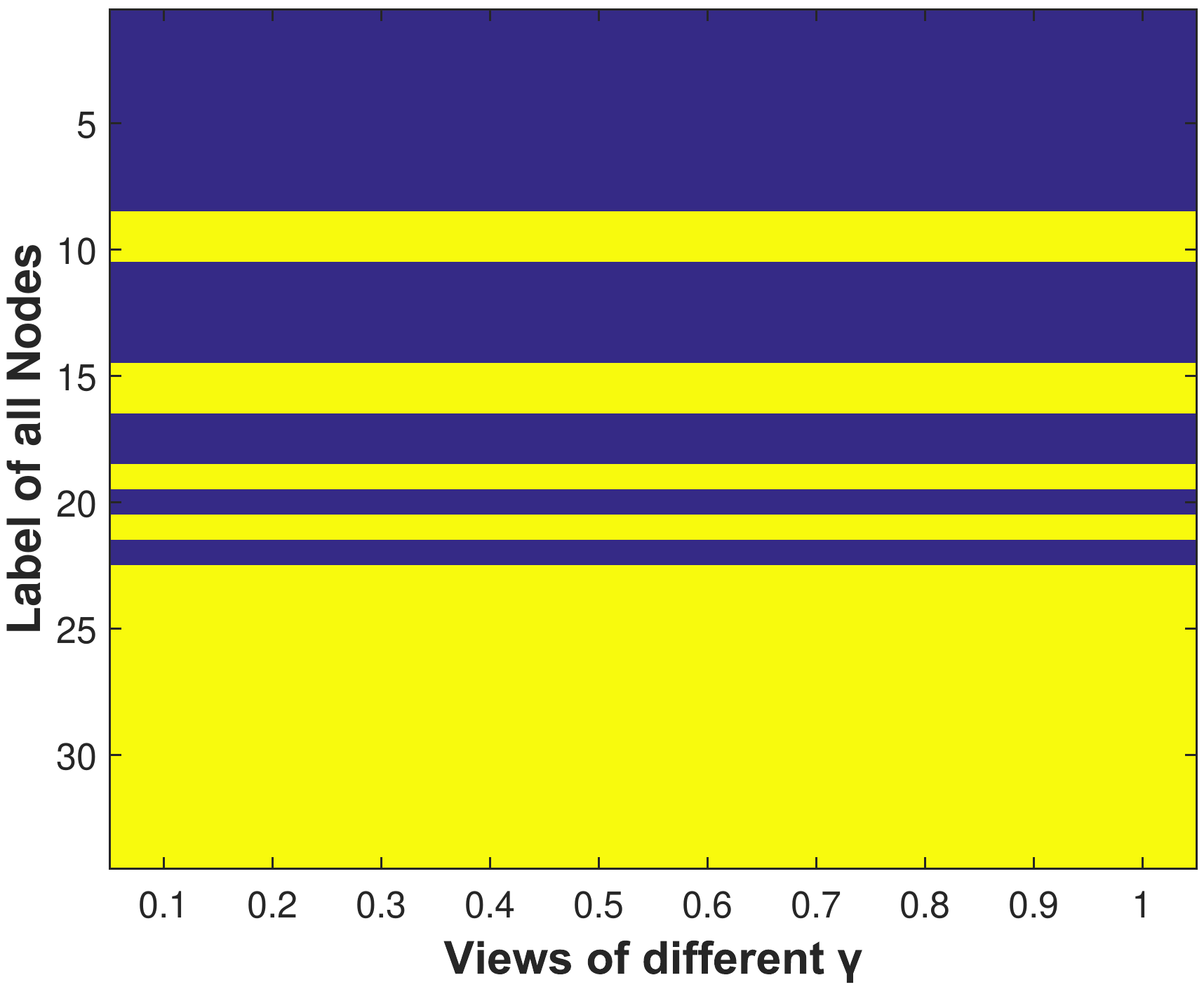}\label{fig:omega10}}%
}
\caption{Community detection results with different parameters. The community assignment is distinguished by different colors. The network consists of ten identical layers each of which is the network of Zachary Karate Club with resolution parameter $\gamma_s \in \{0.1, 0.2, \dots, 1\}$ and detection is performed with coupling strength parameter $\omega = 0, 0.01, 0.1, 1, 10$ respectively.}
\label{fig:parameterAnalysis}
\end{figure*}

\subsection{Parameter Analysis}
In order to study how the parameters (i.e. $\gamma_s$ and $\omega$) in the proposed method influence the community detection results, we conduct experiments with similar experimental settings as \citeauthor{Mucha} do in \cite{mucha:2010community}. We construct a ten-layer network with resolution parameter $\gamma_s \in \{0.1, 0.2, \dots, 1\}$, where the adjacency of each layer is the benchmark network Zachary Karate network\cite{zachary:1977information} and we assume the between-layer coupling exists between any pair of nodes and their copies. We perform community detection on the generated network with different coupling strength parameter $\omega \in \{0, 0.01, 0.1, 1, 10\}$ and the community assignment for each node in ten layers is depicted with different colors.

From Fig. \ref{fig:parameterAnalysis} we can see, when $\omega = 0$, the layers show great divergence due to the value of resolution parameter $\gamma_s$. As $\gamma_s$ grows, the network is inclined to split into subcommunities. By comparing with standard community label, we see the detection result with parameter $\gamma_s$ setting from 0.5 to 0.9 matches the ground truth, while there are misclassification in the rest. As $\omega$ increases, we see the nodes in different layers tend to be assigned to the same community. When $\omega = 1$, we see that every node has the same community label as its copies in other layers, and the detection result consistent with the ground truth.

We can then conclude that, the resolution parameter $\gamma_s$ controls the tendency of the splitting and the coupling strength parameter $\omega$ controls the consistency of the community assignment between layers. Too large or too small $\gamma_s$ will cause misclassification, which can be fixed, however, by the between-layer couplings. Meanwhile, too small $\omega$ will lead to the isolation between layers. When there are noises in the network data, the result can be poor (as shown in Fig. \ref{fig:omega0}) since cross-layer information has not been fully utilized. Nevertheless, the peculiarity of each layer will be damaged by large $\omega$ (as shown in Fig. \ref{fig:omega1}).

In section \ref{sec:comparison} we will compare the performance of several algorithms as the network scale varies, where we bring in a parameter $\rho$ to explicitly control the coupling density. However, since $\rho$ reflects the density of the raw network data, we can consider it as a super parameter that is unalterable once the network is given.

\begin{table*}[!htp]
\begin{center}
\vskip -0.1in
\caption{Comparison of modularity result of CKM-Physicians Innovation network. The best mean values are marked in bold.}
\label{tab:CKM-P}
\begin{tabular}{|c|c|c|c|c|c|c|c|c|c|c|c|c|c|}
\hline
 &$\rho = 0$ &$\rho = .1$ &$\rho = .2$ &$\rho = .3$ &$\rho = .4$ &$\rho = .5$ &$\rho = .6$ &$\rho = .7$ &$\rho = .8$ &$\rho = .9$ &$\rho = 1$ &Variance &Mean \\
\hline
mSpec&1626.5	&1669.1	&1693.3	&1706.5	&1482.5	&1920.8	&1770.6	&1726.9	&2067.8	&2409.9	&2792.2	&295131	&\textbf{2033.4}	\\
\hline
mLouv&431.8	&707.2	&883.5	&1059.7	&1313.3	&1687.7	&1931.3	&2119.0	&2564.3	&2726.9	&3073.7	&940089	&1908.2	\\
\hline
sMeanSpec&427.8	&703.8	&859.8	&1039.8	&1283.8	&1655.8	&1931.8	&2119.8	&2539.8	&2731.8	&3063.8	&947230	&1898.0	\\
\hline
sFullSpec&1622.9	&1848.9	&1773.5	&1753.0	&1838.9	&1964.9	&2008.9	&2020.9	&2080.9	&2104.9	&2152.9	&32522.5	&1964.4	\\
\hline
\end{tabular}
\end{center}
\end{table*}

\begin{table*}[!htp]
\begin{center}
\vskip -0.1in
\caption{Comparison of modularity result of CS-Aarhus network. The best mean values are marked in bold.}
\label{tab:CS-A}
\begin{tabular}{|c|c|c|c|c|c|c|c|c|c|c|c|c|c|}
\hline
 &$\rho = 0$ &$\rho = .1$ &$\rho = .2$ &$\rho = .3$ &$\rho = .4$ &$\rho = .5$ &$\rho = .6$ &$\rho = .7$ &$\rho = .8$ &$\rho = .9$ &$\rho = 1$ &Variance &Mean \\
\hline
mSpec&409.0	&516.2	&523.4	&501.6	&622.4	&544.4	&781.9	&695.6	&915.6	&1078.7	&1378.2	&151061	&\textbf{823.3}	\\
\hline
mLouv&-587.5	&-301.3	&-200.8	&-125.9	&120.5	&408.5	&700.5	&920.5	&1124.5	&1358.7	&1652.5	&663111	&647.6	\\
\hline
sMeanSpec&-692.9	&-400.9	&-293.9	&-216.9	&15.1	&303.1	&595.1	&815.1	&1019.1	&1259.1	&1547.1	&660634	&544.6	\\
\hline
sFullSpec&406.4	&472.4	&468.4	&472.4	&538.4	&568.4	&660.4	&676.4	&726.4	&786.6	&852.4	&25416.1	&639.0	\\
\hline

\end{tabular}
\end{center}
\end{table*}

\begin{table*}[!htp]
\begin{center}
\vskip -0.1in
\caption{Comparison of modularity result of Kapferer Tailor Shop network. The best mean values are marked in bold.}
\label{tab:KTS}
\begin{tabular}{|c|c|c|c|c|c|c|c|c|c|c|c|c|c|}
\hline
 &$\rho = 0$ &$\rho = .1$ &$\rho = .2$ &$\rho = .3$ &$\rho = .4$ &$\rho = .5$ &$\rho = .6$ &$\rho = .7$ &$\rho = .8$ &$\rho = .9$ &$\rho = 1$ &Variance &Mean \\
\hline
mSpec&243.7	&266.7	&291.7	&251.3	&262.1	&271.2	&328.1	&382.6	&480.7	&501.8	&546.4	&22606	&\textbf{384.2}	\\
\hline
mLouv&-173.6	&-84.0	&-30.0	&20.0	&66.9	&176.0	&306.4	&406.4	&536.0	&574.4	&638.4	&98785.2	&293.3	\\
\hline
sMeanSpec&-223.6	&-131.6	&-81.3	&-27.6	&20.4	&128.4	&256.4	&356.4	&488.4	&524.4	&588.4	&98600.0	&244.7	\\
\hline
sFullSpec&243.4	&219.4	&205.4	&261.4	&285.4	&313.4	&319.4	&337.4	&367.4	&395.4	&381.4	&3995.3	&320.8	\\
\hline

\end{tabular}
\end{center}
\end{table*}

\begin{table*}[!htp]
\begin{center}
\vskip -0.1in
\caption{Comparison of modularity result of Krackhardt High-Tech network. The best mean values are marked in bold.}
\label{tab:KHT}
\begin{tabular}{|c|c|c|c|c|c|c|c|c|c|c|c|c|c|}
\hline
 &$\rho = 0$ &$\rho = .1$ &$\rho = .2$ &$\rho = .3$ &$\rho = .4$ &$\rho = .5$ &$\rho = .6$ &$\rho = .7$ &$\rho = .8$ &$\rho = .9$ &$\rho = 1$ &Variance &Mean \\
\hline
mSpec&73.9	&69.6	&65.5086	&65.3	&84.7	&87.6	&99.9	&108.5	&111.3	&125.2	&171.2	&1432.1	&\textbf{106.6}	\\
\hline
mLouv&-48.5	&-24.5	&-10.5	&-4.5	&43.5	&27.5	&75.5	&87.5	&95.5	&135.5	&183.5	&6623.35	&70.1	\\
\hline
sMeanSpec&-58.4	&-34.4	&-23.6	&-14.4	&33.6	&17.6	&65.6	&77.6	&85.6	&125.6	&173.6	&6701.1	&60.5	\\
\hline
sFullSpec&29.0	&49.0	&48.0	&53.0	&83.0	&77.0	&83.0	&91.0	&99.0	&105.0	&131.0	&1163.42	&85.7	\\
\hline
\end{tabular}
\end{center}
\end{table*}

\begin{table*}[!htp]
\begin{center}
\vskip -0.1in
\caption{Comparison of modularity result of London Transportation network. The best mean values are marked in bold.}
\label{tab:LT}
\begin{tabular}{|c|c|c|c|c|c|c|c|c|c|c|c|c|c|}
\hline
 &$\rho = 0$ &$\rho = .1$ &$\rho = .2$ &$\rho = .3$ &$\rho = .4$ &$\rho = .5$ &$\rho = .6$ &$\rho = .7$ &$\rho = .8$ &$\rho = .9$ &$\rho = 1$ &Variance &Mean \\
\hline
mSpec&29.5	&29.5	&29.5	&29.5	&32.9	&34.9	&37.0	&43.5	&37.7	&45.6	&53.5	&102.3	&\textbf{39.4}	\\
\hline
mLouv&-2.2	&1.8	&1.8	&5.8	&17.8	&21.8	&25.8	&41.8	&37.8	&45.8	&53.8	&464.9	&28.3	\\
\hline
sMeanSpec&-2.5	&1.5	&5.5	&5.5	&17.5	&21.5	&25.5	&41.5	&37.5	&45.5	&53.5	&464.9	&28.0	\\
\hline
sFullSpec&24.2	&26.2	&24.2	&24.2	&34.2	&28.2	&36.2	&34.2	&38.2	&36.2	&40.2	&61.5	&33.9	\\
\hline
\end{tabular}
\end{center}
\end{table*}

\begin{table*}[!htp]
\begin{center}
\vskip -0.1in
\caption{Comparison of modularity result of Padgett Florentine Families network. The best mean values are marked in bold.}
\label{tab:PFF}
\begin{tabular}{|c|c|c|c|c|c|c|c|c|c|c|c|c|c|}
\hline
 &$\rho = 0$ &$\rho = .1$ &$\rho = .2$ &$\rho = .3$ &$\rho = .4$ &$\rho = .5$ &$\rho = .6$ &$\rho = .7$ &$\rho = .8$ &$\rho = .9$ &$\rho = 1$ &Variance &Mean \\
\hline
mSpec&176.9	&191.4	&189.9	&201.4	&218.8	&236.1	&228.4	&226.0	&299.1	&313.9	&345.9	&4803.6	&\textbf{257.1}	\\
\hline
mLouv&48.1	&96.1	&106.1	&124.1	&160.1	&216.1	&228.1	&212.1	&300.1	&320.1	&352.1	&12262.7	&222.9	\\
\hline
sMeanSpec&45.0	&93.0	&103.0	&121.0	&157.0	&213.0	&225.0	&209.0	&297.0	&317.0	&349.0	&12262.7	&219.9	\\
\hline
sFullSpec&176.3	&192.3	&194.3	&200.3	&214.3	&238.3	&234.3	&224.3	&236.3	&262.3	&252.2853	&885.8	&227.6	\\
\hline
\end{tabular}
\end{center}
\end{table*}

\begin{table*}[!htp]
\begin{center}
\vskip -0.1in
\caption{Comparison of modularity result of Vickers Class Relation network. The best mean values are marked in bold.}
\label{tab:VCR}
\begin{tabular}{|c|c|c|c|c|c|c|c|c|c|c|c|c|c|}
\hline
 &$\rho = 0$ &$\rho = .1$ &$\rho = .2$ &$\rho = .3$ &$\rho = .4$ &$\rho = .5$ &$\rho = .6$ &$\rho = .7$ &$\rho = .8$ &$\rho = .9$ &$\rho = 1$ &Variance &Mean \\
\hline
mSpec&635.9	&742.7	&787.6	&794.1	&864.4	&1220.6	&1395.1	&1392.2	&1787.7	&2016.4	&2407.6	&541037	&\textbf{1467.7}	\\
\hline
mLouv&-1415.3	&-946.4	&-613.0	&-506.8	&-127.4	&412.2	&929.4	&1124.8	&1785.3	&2113.1	&2612.5	&2.17+06	&817.7	\\
\hline
sMeanSpec&-1494.5	&-1023.2	&-830.1	&-586.5	&-209.9	&326.1	&849.4	&1038.1	&1702.1	&2030.7	&2526.13 &2.16e+06	&735.3	\\
\hline
sFullSpec&546.8	&646.1	&679.3	&631.0	&703.9	&718.1	&798.7	&810.3	&829.6	&932.4	&973.0	&25725.2	&788.1	\\
\hline
\end{tabular}
\end{center}
\vskip -0.1in
\end{table*}

\subsection{Comparison Results}
\label{sec:comparison}

For comparison, several state-of-the-art approaches are used so as to evaluate the performance of the proposed optimization method (mSpec): 1) mLouv: multilayer Louvain-like method plus KL-swap improvement, which is the most widely adopted heuristic method for modularity optimization\cite{mucha:2010community}; 2) sMeanSpec: single-layer spectral optimization method that will be applied on the mean of adjacency matrices of all layers\cite{tang:2010community}; 3) sFullSpec: single-layer spectral optimization method applied on each layer\cite{newman:2004finding,newman:2006modularity}.

In order to examine the reliability of the proposed method, the detection is performed over seven datasets with different between-layer coupling density $\rho$. The parameter $\rho$ depending on the raw network data reflects how closely connected any two layers are, and in experiments we generate random between-layer couplings according to the probability $\rho$. The nodes are linked with all its copies in other layers when $\rho = 1$ and there is no couplings at all when $\rho = 0$.
The result is evaluated by the modularity value $Q$ computed according to Eq. \eqref{eq:ourmetric} using the community assignment of each algorithm, as shown in Tables \ref{tab:CKM-P} to \ref{tab:VCR}.
The variance and mean of the modularity value reflect the stability and reliability of each algorithm against network with different between-layer coupling scales.

As the results suggest, the proposed method significantly outperforms the existing methods, achieving $18.65\%$ improvement over the second best in terms of mean modularity values while maintaining a relatively low variances. The mLouv method and sMeanSpec method show low $Q$ when the couplings are sparse (small $\rho$) and high $Q$ when the couplings are dense (large $\rho$), while sFullSpec performs oppositely. This is because the mLouv and sMeanSpec methods incline to look for a global community label for all nodes and ignore the peculiarity of each layers, so that when the couplings are sparse (which implies high heterogeneity between layers), such algorithm fail to make a distinguished assignment. Similarly, the performance of sFullSpec degenerates seriously when the couplings are dense since it runs detection over each layer respectively and lacks the consideration of consistency. The proposed method is based on a supra-adjacency representation of the multilayer network, with $\omega$ dominates the consistency. This guarantees the reliable performance of the proposed method against networks with different conditions of the connection between layers. In a nutshell, the proposed method performs stably as the coupling density varies so that is relatively reliable when the condition of the raw network is unclear.

\subsection{Case Study: EEG Network}
The event-related potentials (ERPs) which are measured by means of electroencephalography (EEG) is the measured brain response of testee with a specific stimuli\cite{cahn:2006meditation, dietrich:2010review}.
Since the EEG monitoring collects electrical impulse data from the electrodes placed on the scalp, it should be totally noninvasive in most cases except for an inevitable invasive electrode for specific application.
Moreover, the monitoring process is silent so that the auditory disturbance is reduced to a very subtle level and is tolerant to subject movement.
Owing to the numerous advantages, EEG is widely adopted as the analysis tool for brain activity, especially on children testees.
Nevertheless, the traditional output of the EEG monitoring manifests as waveforms, so that the analysis of them is unintuitive and usually relies on the experiential judgements of the EEG providers.
In recent years, more and more research focus is concentrated on the analysis of EEG data, but almost all of such work focuses on the average performance of similar testees, which may lead to the loss of information about each distinct testee\cite{alexander:2012discovery, chen:2008revealing, meunier:2009hierarchical}.
In this experiment, we attempt to apply the proposed method on the signed multilayer network generated from the EEG data to explore the functional performance of the regions of brain.
We compare the detected result with a standard empirical brain functional region division to find a surprising match between clinical experience and graph data mining\cite{power:2011functional}.

\begin{figure*}[!htp]
\centerline{
\subfigure[testee 1]{\includegraphics[width=0.26\linewidth]{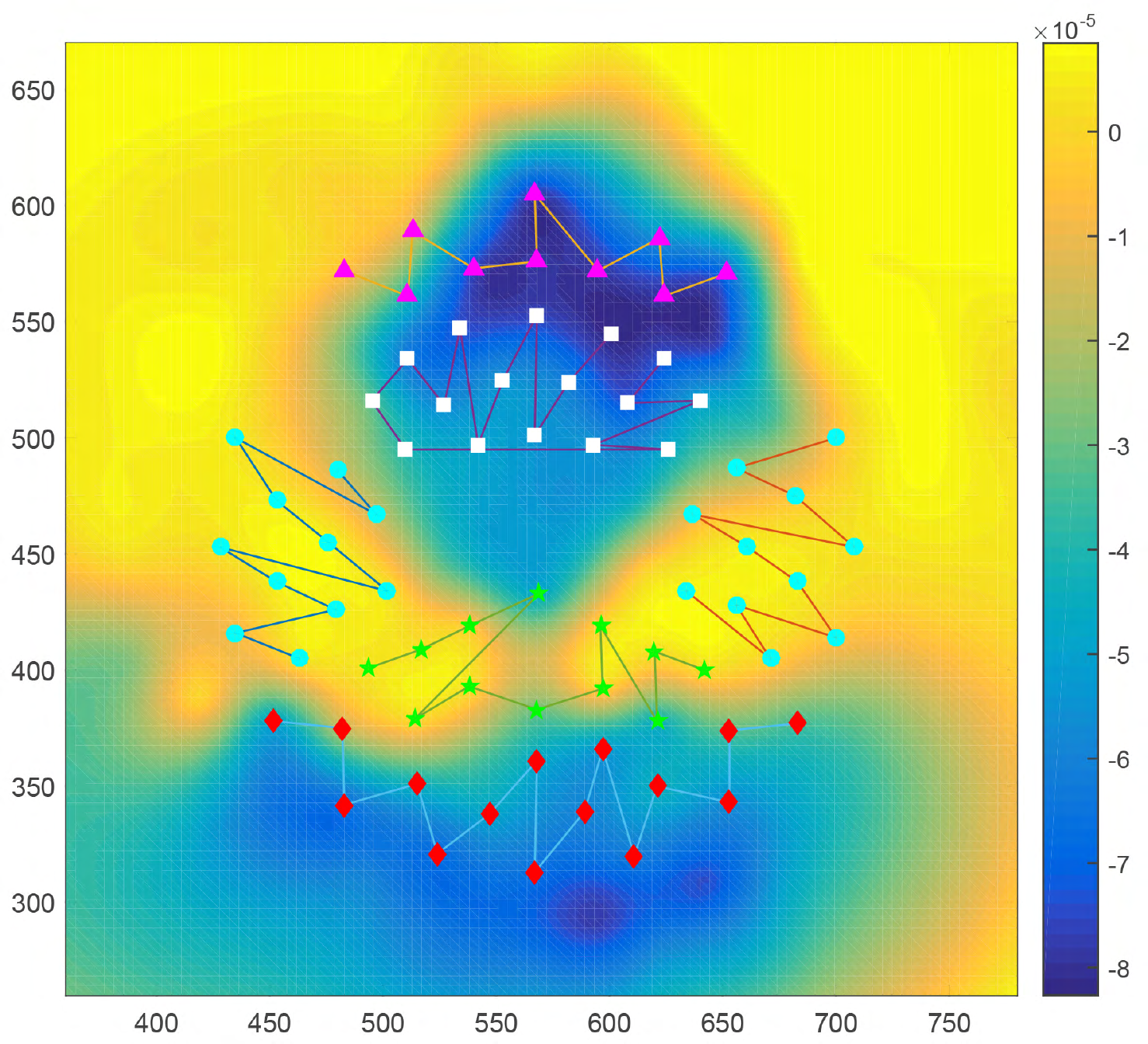}\label{fig:person1}}%
\subfigure[testee 2]{\includegraphics[width=0.27\linewidth]{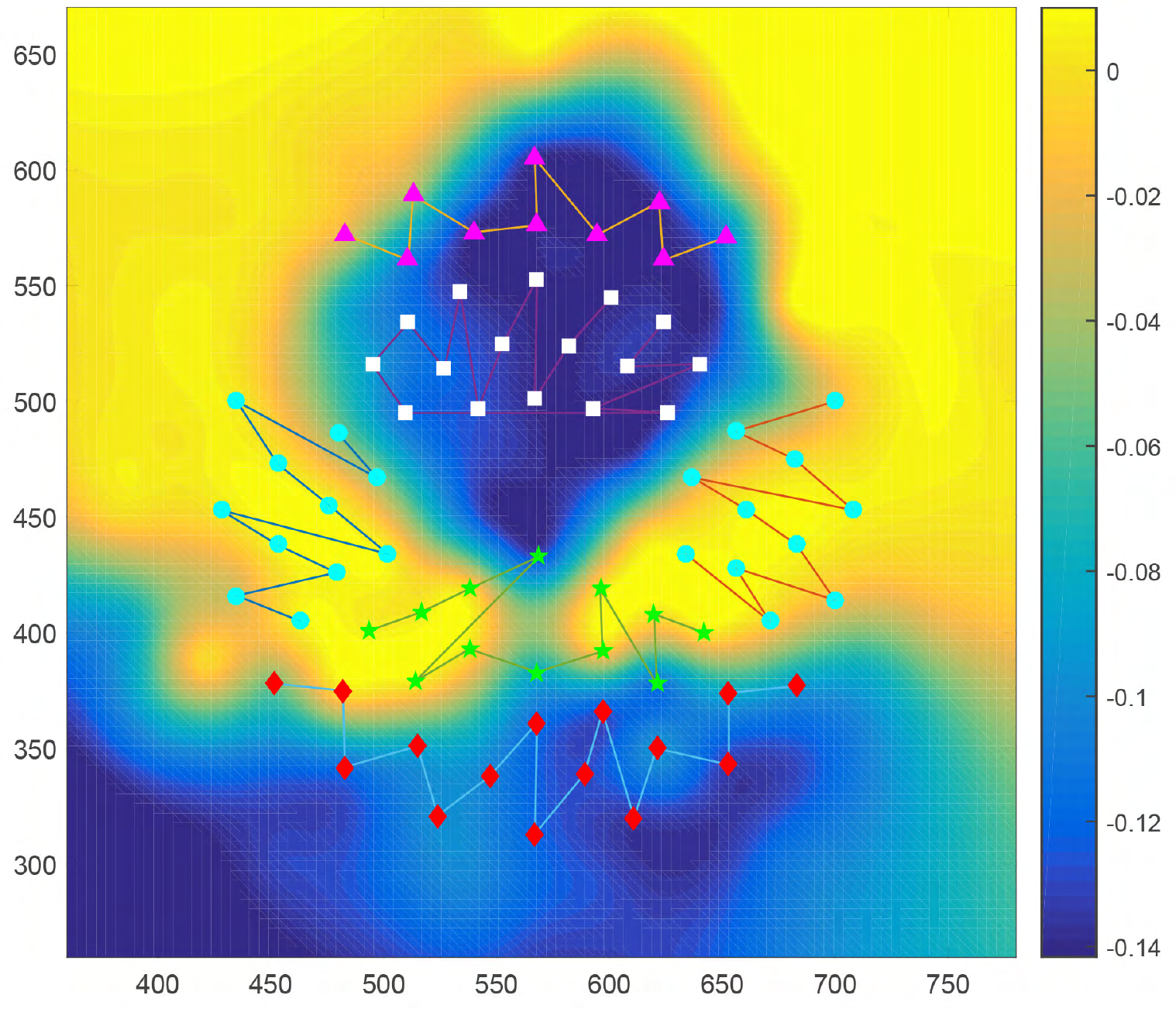}\label{fig:person2}}%
\subfigure[testee 3]{\includegraphics[width=0.26\linewidth]{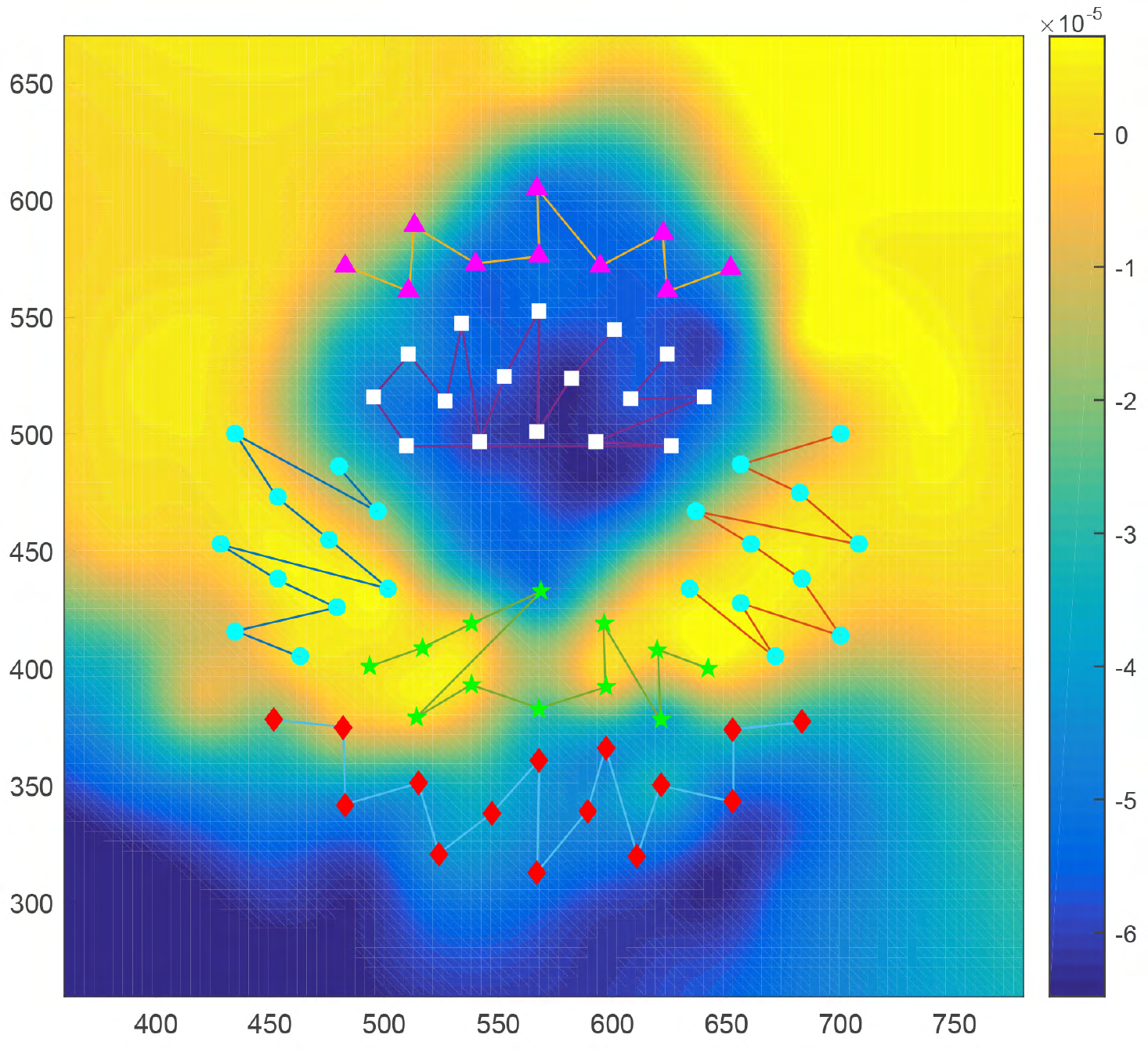}\label{fig:person3}}%
\subfigure[testee 4]{\includegraphics[width=0.26\linewidth]{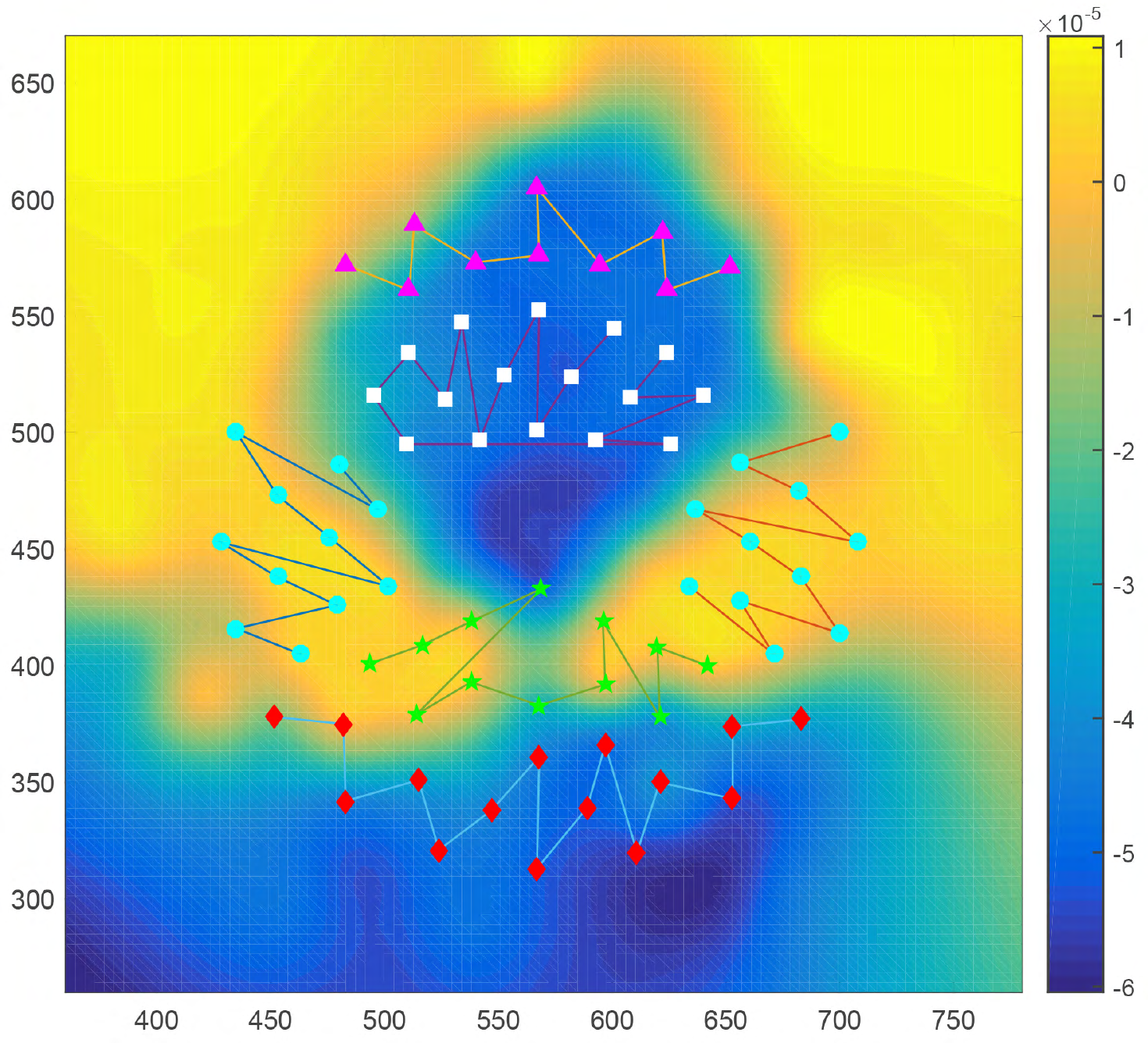}\label{fig:person4}}%
}
\caption{The detection result of EEG network. We randomly pick 4 layers from the multilayer network. The standard brain region division is plotted with different symbols: i) purple triangle: prefrontal cortex that controls thinking, perception, information memory and attention; ii) white square: premotor cortex that controls eye movements; iii) blue circle: auditory cortex that controls the audition; iv) green star: somatosensory cortex that controls the sense of touch; v) red diamond: visual cortex that controls the sense of sight. The detected result is presented as the topographic map of the brain where we directly treat the dominant eigenvector $\mu$ as the community label. The blue region corresponds to the negative terms of $\mu$ while the yellow region corresponds to the positive terms, where the darkness indicates the magnitude of corresponding label value.}
\label{fig:EEG}
\end{figure*}

We regard the 128 electrodes and a standard control electrode placed on the testee's scalp as 129 nodes involved, and calculate the correlation coefficients between the ERPs recorded from each pair of electrodes when the testee is given a series of visual stimuli as the edge weights between them.
Thus we generate a single-layer network based on the EEG data of one test record of a specific testee.
By combining the networks generated in this way from 11 testees and their several test records, we obtain a two-aspect EEG network that contains the information of the brain activities of all testees.
Since the electrodes are placed identically for every testee, we assume the between-layer coupling exists between each pair of corresponding electrodes.
We can adjust the parameter $\gamma_s$ to control the resolution and $\omega$ to control the consistency of the detection result of each testee.
The detection results on the first four testees are shown in Fig. \ref{fig:EEG}.

We find that the EEG networks are always divided into two communities, yellow and blue, in all experiments.
By comparing the detection results with the corresponding adjacency, we observe that the edges with negative weights mainly lie between the two communities and within each community the nodes are connected by the edges with positive weights.
Therefore, in order to better illustrate the brain terrain, we directly treat the dominant eigenvector $\mathbf{u}_M$ of the modularity matrix as the detected community labels of the corresponding nodes for plotting since such non-binary labels make it possible to picture the contour of the brain. Say, the dominant eigenvector is $(0.5, 0.2, -0.1, -1)$ and the label vector will also be $(0.5, 0.2, -0.1, -1)$, where the last two nodes will be dyed blue (the darkness distinguishes the magnitude) and the first two will be dyed yellow.
Meanwhile, since such treatment also maximize the modularity function, the result is more accurate and reliable than discrete community labels.
We present the continuous community label as the topographic map of the brain where the two communities correspond to regions with different color.
By adding the standard brain function region division to the figures, we find the detection results reach a surprising match with the widely accepted brain functional partition.
The visual cortex (red diamond), prefrontal cortex (purple triangle) and the premotor cortex (white square) share the same community while the auditory cortex which is denoted with blue circles belongs to the other community.
The former is more or less relevant to the visual and attention while the latter is closely related with audition.
The results coincide with the clinical experience that the visual and audition always demonstrate relatively strong divergence and interaction.
Moreover, from the colorbar attached we can notice the magnitude of continuous community label of the blue part which corresponds to visual brain region is much higher than that of the yellow part which refers to the auditory region.
The magnitude of the continuous label indicates the contribution of the node to the global modularity value, which can imply how active the region is during the test.
Therefore, we see the visual region is much more active than the auditory region, which coincides our intuition.

To sum up, this experiment on EEG network shows encouraging results about the feasibility of the proposed method on empirical networks. It also provides a new direction of the application of the proposed method and similar approaches.

\section{Conclusion}
\label{sec:conclusion}
In this paper, we discussed the representation of multilayer networks with multiple aspects and then derived the multilayer modularity based on the assumption of the contribution of the edges and couplings.
According to the derivation, we demonstrate that the modularity prefers the community structure where the edges and couplings are densely distributed within the communities.
Then we proposed a spectral bisection method for optimization of the modularity based on the supra-adjacency representation.
In the experiment section, we reported the performance of the proposed evaluation metric as the parameters change and the comparison result with some other baseline methods.
We applied the proposed method on a two-aspect EEG network as an attempt of application, and the results tend to coincide with the functional region of the brain.

\section*{Acknowledgements}
We'd like to thank Sun Yat-sen Memorial Hospital for providing EEG data. This work was supported by NSFC (No. 61502543 and No. 61573387
), GuangZhou Program (No. 201508010032), Guangdong Natural Science Funds for Distinguished Young Scholar (16050000051), the PhD Start-up Fund of Natural Science Foundation of Guangdong Province, China (2014A030310180), the Fundamental Research Funds for the Central Universities (16lgzd15), NSF through grants III-1526499, and CNS-1115234.

\ifCLASSOPTIONcaptionsoff
  \newpage
\fi
 {
\bibliographystyle{IEEEtran}
\bibliography{TNSE}
}

\end{document}